\documentclass[prd,aps,nofootinbib,onecolumn]{revtex4}
\usepackage{amsmath}
\usepackage{epsfig}
\usepackage{amssymb}
\usepackage{color}

\begin{document}

\title{Nonleptonic two-body decays of charmed mesons}
\author{Fu-Sheng Yu, Xiao-Xia Wang, and Cai-Dian L\"{u}}
\affiliation{Institute of High Energy Physics and Theoretical
Physics Center for Science Facilities, Chinese Academy of Sciences,
Beijing 100049, People's Republic of China}

\begin{abstract}
Nonleptonic decays of charmed mesons into two pseudoscalar mesons or
one pseudoscalar meson and one vector meson are studied on the basis
of a generalized factorization method considering the resonance
effects in the pole model for the annihilation contributions. Large
strong phases between different topological diagrams are considered
in this work, simply taking the phase in the coefficients $a_i$. We
find that the annihilation-type contributions calculated in the pole
model are large in both of the $PP$ and $PV$ modes, which make our
numerical results agree with the experimental data better than those
previous calculations.
\end{abstract}

%%%%%%%%%%%%%%%%%%%%%%%%%%%%%%%%%%%%%%%%%%%%%%%%%%%%%%%%%%%%%%%%%%%%%%%%%%%%%%%%%%%%%%%%%%%%%%%%%%%%%%%%%%%%%%%%%%%%%
\maketitle


\section{Introduction}
Nonleptonic decays of charmed mesons are very interesting as they
can provide useful information on flavor mixing, CP violation, and
strong interactions\cite{Artuso:2008vf}. They may also shed light on
any new physics signal through $D^0-\bar{D}^0$ mixing and rare
decays\cite{raredecay,rarefajfer,cpv,daonenggao,yabsley,lipkin}. The
CLEO-c and two B factories experiments have given many new results
on this subject and more are expected soon from the BES-III
experiment. Besides, theoretical studies have been in progress for
decades.

With the heavy quark effective theory, many QCD-inspired approaches,
such as the QCD factorization approach \cite{qcdf}, the perturbative
QCD approach \cite{pqcd}, and the soft-collinear effective theory
\cite{scet}, successfully describe the hadronic B decays. However,
this is not the case for the $D$ meson decays. These approaches do
not work well here, due to the mass of charm quark, of order 1.5GeV,
which is not heavy enough for a sensible heavy quark expansion,
neither light enough to apply the chiral perturbation theory.

After decades of studies, the  factorization approach is still one
of the effective ways to deal with the  two-body charmed meson
decays\cite{BSW}. However, it is well known that some difficulties
exist in the naive factorization approach: the Wilson coefficients
$a_1(\mu)$ and $a_2(\mu)$ of effective operators are renormalization
scale and $\gamma_5$-scheme dependent; and the color-suppressed
processes are not calculated well due to the smallness of $a_2$,
etc. In order to solve these problems, the so-called generalized
factorization approach was proposed
\cite{generalizedfactorizationscaleindependent}. The Wilson
coefficients $a_1(\mu)$ and $a_2(\mu)$ are not from direct
calculation any more but are effective  coefficients  to accommodate
the important nonfactorizable corrections
\cite{generalizedfactorizationnonfactorizableterm}. In these naive
or  generalized factorization approaches,  there is almost no strong
phase. But large strong phases have been found  from $D$ decay
experiments. A corresponding large relative strong phase between the
factorization coefficients $a_1$ and $a_2$ has been discussed in
\cite{a1a2phasehaiyangcheng,a1a2phaseyufengzhou}.

Unsatisfied by the short comings of the factorization approach, the
model-independent diagrammatic approach, with various topological
amplitudes extracted from the data, is recently applied to two-body
hadronic $D$ decays
 \cite{Rosner:1999xd,diagrammatic,diagram,diagrammaticchengchiang}.  They use SU(3) symmetry
 in their analysis to avoid model calculations. All the parameters are fitted from experiments to
 give a better agreement with the experimental data but with less predictive power. More precise predictions are limited in
 this approach due to the uncontrolled SU(3) breaking effect\cite{a1a2phaseyufengzhou}. These
 analyses also show that large
  annihilation type
contributions  are needed to explain the data, which can not be
calculated in the naive or generalized factorization approaches. The
kind of pure annihilation type $D$ meson decays also needs to be
systematically analyzed \cite{du}.

In another aspect, the hadronic picture description of nonleptonic
weak decays has a longer history, because of their nonperturbative
feature. Based on the idea of vector dominance, which is discussed
on strange particle decays\cite{vectordominance}, the pole-dominance
model of two-body nonleptonic decays is proposed \cite{polemodel}.
Beyond the vector dominance pole, this model also involves scalar,
pseudoscalar, and axial-vector poles. For simplicity, only the
lowest-lying poles are considered. This model has already been
applied to charmed meson and bottom meson decays
\cite{polemodel,polemodelofLuCaiDian}, where it is approved that
this model is more or less equivalent to the factorization approach
in the first order approximation.

In this work, we will use the generalized factorization approach but
with a relative strong phase between the Wilson coefficients $a_1$
and $a_2$ to accommodate the nonfactorizable contributions. For the
uncalculable annihilation type contributions, we use  the pole
model. In this case, we can really calculate most of the important
contributions in the hadronic $D$ decays, which are demonstrated in
the model-independent diagrammatic analysis
\cite{diagrammaticchengchiang}.

The outline of this paper is as follows. In Sec.II, we give the
formulas of this work, showing the generalized factorization
approach and the pole-dominance model. In Sec.III, our results are
given and compared to the experimental data and those of the
diagrammatic approach and the calculations considering the
final-state interaction of nearby resonances effects. Summary and
conclusions are followed.

%%%%%%%%%%%%%%%%%%%%%%%%%%%%%%%%%%%%%%%%%%%%%%%%%%%%%%%%%%%%%%%%%%%%%%%%%%%%%%%%%%%%%%%%%%%%%%%%%%%%%%%%%%%%%%%%

\section{Formalism}

\subsection{The factorization approach}
First we begin with the weak effective Hamiltonian $H_{eff}$ for the
$\Delta C=1$ transition \cite{buras}:
\begin{equation}\label{effectiveHamiltonian}
\mathcal{H}_{eff}=\frac{G_F}{\sqrt{2}}V_{CKM}(C_1O_1+C_2O_2)+h.c.,
\end{equation}
where $V_{CKM}$ is the corresponding Cabibbo-Kobayashi- Maskawa
(CKM) matrix elements, $C_{1,2}$ are the Wilson coefficients. The
current-current operators $O_{1,2}$ are
\begin{eqnarray}
O_1=\bar{u}_{\alpha}\gamma_{\mu}(1-\gamma_5)q_{2\beta}\cdot\bar{q}_{3\beta}\gamma^{\mu}(1-\gamma_5)c_{\alpha},
\nonumber\\
O_2=\bar{u}_{\alpha}\gamma_{\mu}(1-\gamma_5)q_{2\alpha}\cdot\bar{q}_{3\beta}\gamma^{\mu}(1-\gamma_5)c_{\beta},
\end{eqnarray}
where $\alpha,\beta$ are color indices, and $q_{2,3}$ are  $d$ or
$s$ quarks.

%%%%%%%%%%%%%%%%%%%%%%%%%%%%%%%%%%%%%%%%%%%%%%%%%%%%%
\begin{figure}
\begin{center}
\includegraphics[scale=0.5]{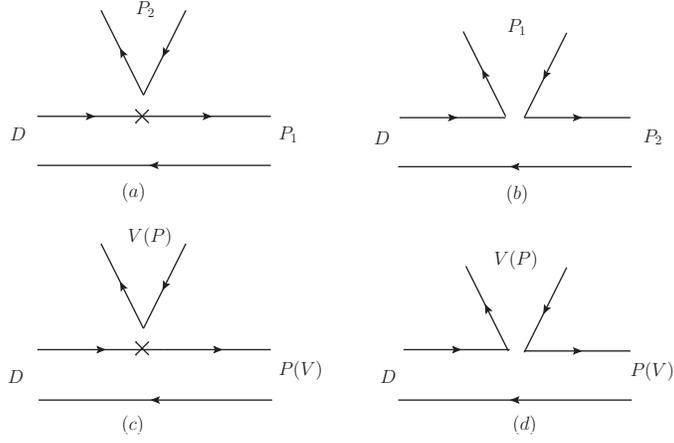}
\end{center}
\caption{Emission-type diagrams in the factorization
approach.}\label{FIGfac}
\end{figure}
%%%%%%%%%%%%%%%%%%%%%%%%%%%%%%%%%%%%%%%%%%%%%%%%%%%%%

The  color-favored emission diagram corresponding to $D\to PP $
decays,  with $P$  representing a pseudoscalar meson, is shown in
Fig. \ref{FIGfac} (a). Under the factorization hypothesis, the
transition matrix element of hadronic two-body charmed meson decays
is factorized into two parts \cite{BSW}:
\begin{eqnarray}
\langle P_1P_2|\mathcal{H}_{eff}|D \rangle
&=&\frac{G_F}{\sqrt{2}}V_{CKM}a_1\langle P_2|\bar{u}
\gamma_{\mu}(1-\gamma_5)q_{2}|0\rangle \langle
P_1|\bar{q}_{3}\gamma^{\mu}(1-\gamma_5)c|D\rangle
.\label{matrixelement}
\end{eqnarray}
Similarly, the contribution of the color-suppressed diagram shown in
Fig. \ref{FIGfac} (b) is given as
\begin{eqnarray}
\langle P_1P_2|\mathcal{H}_{eff}|D \rangle &=&
\frac{G_F}{\sqrt{2}}V_{CKM}a_2\langle P_1|\bar {q}_{3}
\gamma_{\mu}(1-\gamma_5)q_{2}|0\rangle \langle
P_2|\bar{u}\gamma^{\mu}(1-\gamma_5)c|D\rangle,\label{matrixelement2}
\end{eqnarray}
where
\begin{eqnarray}
a_1(\mu)=C_2(\mu)+\frac{C_1(\mu)}{N_c},&&
a_2(\mu)=C_1(\mu)+\frac{C_2(\mu)}{N_c},
\end{eqnarray}
correspond to the color-favored tree diagram ($\mathcal{T}$) and
 the color-suppressed diagram ($\mathcal{C}$) respectively in the naive
factorization, with the number of colors $N_c=3$. They are assumed
to be universal and process-independent in the native factorization
approach. The current matrix elements in
Eqs.(\ref{matrixelement},\ref{matrixelement2}) are evaluated in
terms of transition form factors and decay constants. For $D\to PP$
decays, the form factor is defined as follows:
\begin{eqnarray}
\langle
P(k)|\bar{q}_{3}\gamma_{\mu}(1-\gamma_5)c|D(p)\rangle&=&\bigg[(p+k)_{\mu}-\frac{m_D^2-m_P^2}{q^2}q_{\mu}\bigg]F_1^{D\to
P}(q^2)\nonumber\\
&&+\frac{m_D^2-m_P^2}{q^2}q_{\mu}F_0^{D\to P}(q^2),
\end{eqnarray}
where $q=p-k$, and $F_i$  are the  corresponding transition form
factors. The decay constants $f_P$   of pseudoscalar  mesons are
defined as
\begin{eqnarray}
\langle P(q)|\bar{q}_{1}\gamma_{\mu}(1-\gamma_5)q_{2}|0\rangle&=&i
f_P q_{\mu}.
\end{eqnarray}

In terms of decay constant and transition form factors, the decay
amplitude of Figs.\ref{FIGfac} (a) and (b) are then
\begin{eqnarray}
\langle P_1P_2|\mathcal{H}_{eff}|D\rangle_T
&=&i\frac{G_F}{\sqrt{2}}V_{CKM}
a_{1}f_{P_2}(m_D^2-m_{P_1}^2)F_0^{D\to P_1}(m_{P_2}^2),
\end{eqnarray}
\begin{eqnarray}
\langle P_1P_2|\mathcal{H}_{eff}|D\rangle_C
&=&i\frac{G_F}{\sqrt{2}}V_{CKM}
a_{2}f_{P_1}(m_D^2-m_{P_2}^2)F_0^{D\to P_2}(m_{P_1}^2).
\end{eqnarray}

The diagrams for $D\to PV$ decays, with $V$ denoting a vector meson,
are shown in Figs. \ref{FIGfac} (c) and (d). If the out-emitted
particle is a pseudoscalar meson, the matrix element of Fig.
\ref{FIGfac} (c) is
\begin{eqnarray}
\langle PV|\mathcal{H}_{eff}|D\rangle &=&\frac{G_F}{\sqrt{2}}V_{CKM}
a_{1}\langle P|\bar{u}\gamma_{\mu}(1-\gamma_5)q_2|0\rangle\langle
V|\bar{q}_3\gamma^{\mu}(1-\gamma_5)c|D\rangle.
\label{factorizationDtoV}
\end{eqnarray}
The $D\to V$ transition form factors are usually defined as
\begin{eqnarray}
\langle V(k)|\bar{q}_{3}\gamma_{\mu}(1-\gamma_5)c|D(p)\rangle&=&
\frac{2}{m_D+m_V}\epsilon_{\mu\nu\rho\sigma}\varepsilon^{*\nu}p^{\rho}k^{\sigma}V(q^2)
\nonumber\\
&&-i\bigg(\varepsilon^*_{\mu}-\frac{\varepsilon^*\cdot
q}{q^2}q_{\mu}\bigg)(m_D+m_V)A_1^{D\to V}(q^2) \nonumber\\
&&+i\bigg((p+k)_{\mu}-\frac{m_D^2-m_V^2}{q^2}q_{\mu}\bigg)\frac{\varepsilon^*\cdot
q}{m_D+m_V}A_2^{D\to V}(q^2) \nonumber\\
&&-i\frac{2m_V(\varepsilon^*\cdot q)}{q^2}q_{\mu}A_0^{D\to V}(q^2),
\end{eqnarray}
where $\varepsilon^*$ is the polarization vector of the vector
meson, and $A_i$ and $V$ are corresponding transition form factors.
Utilizing the form factor definitions we get the result for
Eq.(\ref{factorizationDtoV}):
\begin{eqnarray} \langle PV|\mathcal{H}_{eff}|D\rangle
&=& \frac{G_F}{\sqrt{2}}V_{CKM} a_{1}f_Pm_V A_0^{D\to
V}(m_P^2)2(\varepsilon^*\cdot p_D).\label{factorizationDtoV2}
\end{eqnarray}
If a vector meson is out emitted, the matrix element of Fig.
\ref{FIGfac}(c) is
\begin{eqnarray}
\langle PV|\mathcal{H}_{eff}|D\rangle&=&\frac{G_F}{\sqrt{2}}V_{CKM}
a_{1}\langle V|\bar{u}\gamma^{\mu}(1-\gamma_5)q_2|0\rangle\langle
P|\bar{q}_3\gamma^{\mu}(1-\gamma_5)c|D\rangle
\nonumber\\
&=&\frac{G_F}{\sqrt{2}}V_{CKM} a_{1}f_V m_V F_1^{D\to P}(m_V^2) 2
 (\varepsilon^*\cdot p_D),\label{factorizationDtoP}
\end{eqnarray}
where the decay constants   $f_V$ of    vector mesons are defined as
\begin{eqnarray}
\langle V(q)|\bar{q}_{1}\gamma_{\mu}(1-\gamma_5)q_{2}|0\rangle&=&f_V
m_V \varepsilon^*_{\mu}(q).
\end{eqnarray}

For the color-suppressed diagram, Fig.\ref{FIGfac}(d), we have
similar formulas  as
Eqs.(\ref{factorizationDtoV2},\ref{factorizationDtoP}), but with the
Wilson coefficient changed from $a_1$ to $a_2$.

As mentioned in the Introduction, the Wilson coefficients $a_1$ and
$a_2$ are renormalization-scale dependent in the  naive
factorization approach and it fails to describe  the
color-suppressed processes with too small $a_2\approx -0.1$. So the
generalized factorization method is proposed to include the
nonfactorizable contributions
\cite{generalizedfactorizationnonfactorizableterm},
\begin{eqnarray}\label{effa}
a_1^{\rm
eff}=C_2(\mu)+C_1(\mu)\bigg(\frac{1}{N_c}+\chi_1(\mu)\bigg),&&
a_2^{\rm
eff}=C_1(\mu)+C_2(\mu)\bigg(\frac{1}{N_c}+\chi_2(\mu)\bigg),
\end{eqnarray}
where the  terms $\chi_i$ characterize the nonfactorizable
corrections involving vertex corrections, hard spectator
interactions, final-state interactions, resonance effects, etc.
These $\chi_i(\mu)$ will compensate the scale- and scheme-dependence
of the Wilson coefficients, so that $a_i$'s are physical now.
Without confusion, we will drop the superscript "eff" in the
effective Wilson coefficients for convenience in the following
discussions. In the large-$N_c$ approach, the 1/$N_c$ terms are
discarded\cite{largeNc}, equally with a universal nonfactorizable
term $\chi_1=\chi_2=-1/N_c$, hence,
\begin{eqnarray}
a_1\approx C_2(m_c)=1.274, && a_2\approx C_1(m_c)=-0.529.
\end{eqnarray}
This implies a null relative strong phase between the two kinds of
contributions. However, the experimental data tell us that there
should be a large strong phase between $a_1$ and $a_2$. On the other
hand, the existence of relative phases is reasonable for the
importance of inelastic final-state interactions of the $D$ meson
decays, in which the  on-shell intermediate states contribute
imaginary parts. Therefore, we consider a relative phase between the
coefficients $a_1$ and $a_2$ in this work, so that
\begin{eqnarray}
a_1=|a_1|,&& a_2=|a_2|e^{i\delta},
\end{eqnarray}
where we set $a_1$ real for convenience.

\subsection{Pole-dominance Model}

\begin{figure}
\begin{center}
\includegraphics[scale=0.7]{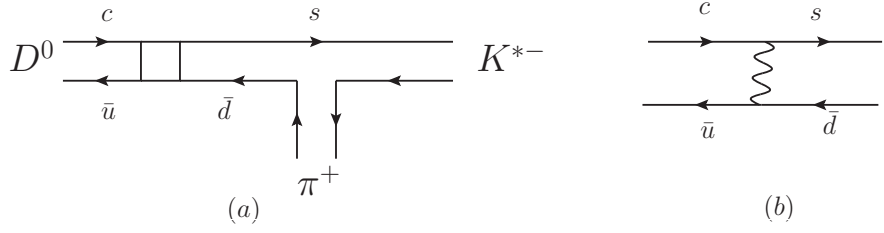}
\end{center}
\caption{Annihilation diagrams in the pole model.}\label{pole1}
\end{figure}

The annihilation type diagrams are neglected as an approximation in
the factorization model. However, considerable contributions come
from  the weak annihilation diagrams in the $D$ decays, which can be
demonstrated by the difference of lifetime between $D^0$ and $D^+$.
Hence, we will calculate them in a single pole-dominance model. For
simplicity, only the lowest-lying poles are considered in the
single-pole model. Taking $D^0\to\pi^+{K}^{*-}$ as an example, the
annihilation-type diagram in the pole model is shown in Fig.
\ref{pole1}(a). $D^0$ goes into $\bar{K}^0$ via  the weak
interaction in Eq. (\ref{effectiveHamiltonian}) shown in terms of
quark lines in Fig. \ref{pole1}(b), and then decays into
$\pi^+K^{*-}$ through the strong interaction. Angular momentum
should be conserved at the weak vertex and all conservation laws be
preserved at the strong vertex. So it is  a pseudoscalar meson as a
resonant state for $D\to PV$ decays. The weak matrix element is
evaluated in the vacuum insertion
approximation\cite{polemodelofLuCaiDian},
\begin{eqnarray}
\langle
\bar{K}^0|\mathcal{H}_{eff}|D^0\rangle&=&\frac{G_F}{\sqrt{2}}V_{cs}^*V_{ud}a_E^{PV}\langle
\bar{K}^0|\bar{s}
\gamma_{\mu}(1-\gamma_5)d|0\rangle\langle0|\bar{u}\gamma^{\mu}(1-\gamma_5)c|D^0\rangle
\nonumber\\
&=&\frac{G_F}{\sqrt{2}}V_{cs}^*V_{ud}a_E^{PV}f_{K}f_{D}m_D^2.
\end{eqnarray}
where the subscript $E$ of the Wilson coefficient $a_E^{PV}$ denotes
 a W-exchange diagram for $D\to PV$, otherwise $a_A^{PV}$
corresponds to W-annihilation contributions. In fact, the effective
Wilson coefficients of the W-annihilation diagrams and W-exchange
diagrams have the same form as $a_1$ and $a_2$ in Eq.(\ref{effa}),
that is
 \begin{eqnarray}
a_E=C_1(\mu)+C_2(\mu)\left({1\over N_c}+\chi_E(\mu)\right),~~~
a_A=C_2(\mu)+C_1(\mu)\left({1\over N_c}+\chi_A(\mu)\right),
 \end{eqnarray}
where $\chi_{A(E)}$ represents the nonfactorizable contributions in
the annihilation (exchange) process. Since the nonfactorizable
contributions in these kinds of diagrams are large and  with
relatively different strong phases, we use different symbols to
avoid confusing in our approach for these collective effective
Wilson coefficients as $a_E$ and $a_A$, respectively. Strong phases
relative to  the emission diagrams are considered in the Wilson
coefficients. The effective strong coupling constant of $\bar{K}^0$
to $\pi^+K^{*-}$ is defined through the Lagrangian
\begin{equation}
\mathcal{L}_{VPP}=ig_{VPP}V^{\mu}(P\overleftrightarrow{\partial}_{\mu}P),
\end{equation}
where $g_{VPP}$ is dimensionless.  Inserting the propagator of the
intermediate $\bar{K}^0$ meson, the decay amplitude is
\begin{eqnarray}
\langle\pi^+K^{*-}|\mathcal{H}_{eff}|D^0\rangle&=&\frac{G_F}{\sqrt{2}}V_{cs}^*V_{ud}a_E^{PV}f_{K}f_{D}g_{K^*K\pi}\frac{m_D^2}{m_D^2-m_{K}^2}2(\varepsilon^*\cdot
p_D).
\end{eqnarray}

%%%%%%%%%%%%%%%%%%%%%%%%%%%%%%%%%%%%%%%%%%%%%%%%%%%%%
%%%%%%%%%%%%%%%%%%%%%%%%%%%%%%%%%%%%%%%%%%%%%%%%%%%%%

Similarly, in $D\to PP$ decays,  it is a scalar meson as a resonant
state. The effective strong coupling constant is described by
\begin{equation}
\mathcal{L}_{SPP}=-g_{SPP}m_{S}SPP,
\end{equation}
where $m_S$ is the mass of the scalar meson.
 Besides, the scalar meson
 decay constant of the vector current is defined as
\begin{eqnarray}
\langle S(p)|\bar{q_2}\gamma_{\mu}q_1|0\rangle=f_Sp_{\mu}.
%&&\langle S|\bar{q_2}q_1|0\rangle=m_S\bar{f}_S,\label{fs1}
\end{eqnarray}

Therefore, the corresponding matrix element is
\begin{eqnarray}
\langle
PP|\mathcal{H}_{eff}|D\rangle&=&-i\frac{G_F}{\sqrt{2}}V_{CKM}a_{A}(a_E)f_{S}f_{D}g_{SPP}\frac{m_D^2m_{S}}{m_D^2-m_{S}^2}.
\end{eqnarray}

As a convenience of reference, the various decay formulas of
individual decay modes are collected in the appendix. Note that some
of the intermediate resonances are unstable particles which have
large width and therefore contribute large relative phases. These
phases are absorbed in the effective Wilson coefficients $a_A$ and
$a_E$ for convenience.

%%%%%%%%%%%%%%%%%%%%%%%%%%%%%%%%%%%%%%%%%%%%%%%%%%%%%%%%%%%%%%%%%%%%%%%%%%%%%%%%%%%%%%%%%%%%%%%%%%%%%%%%%%%%%%%%%%%

\section{Numerical Analysis}

\subsection{Input Parameters}

In order to calculate the emission type diagrams in the
factorization approach, we need to know the transition form factors
and meson decay constants. The decay constants  of $\pi$, $K$, $D$
and $D_s$ are taken from  the particle data group (PDG)
\cite{PDG2010}, others are from \cite{decayconstantsofVectors}, all
of which  are summarized in Table~\ref{decayconstant}. There exist
many models to parametrize the  transition form factors and their
$q^2$ dependence
\cite{formfactorStech,formfactorCLEOc,ffhongweike,formfactorwangwei,formfactorfajfer,ff1,ff2,ff3,ff4,ff5,ff6,ff7,ff8,ff9,ff10,ff11,ff12,ff13,ff14}.
In this work we shall use the dipole model \cite{formfactorStech}:
\begin{eqnarray}\label{formfactoralpha1alpha2}
F(q^2)=\frac{F(0)}{(1-\alpha_1\frac{q^2}{m_{pole}^2}+\alpha_2\frac{q^4}{m_{pole}^4})},
\end{eqnarray}
where $m_{pole}$ is the mass of the pole. The corresponding poles
are $D^*$ for $F_{0,1}^{D\pi,D\eta^{(')},D_sK}$, $D^*_s$ for
$F_{0,1}^{DK,D_s\eta^{(')}}$, $D$ for $A_0^{D\rho,D\omega,D_sK^*}$,
and $D_s$ for $A_0^{DK^*,D_s\phi}$. The transition form factors and
$\alpha_i$ parameters  of $D$ to $\pi$ and $K$ are taken from the
recent CLEO-c measurement \cite{formfactorCLEOc}, $D\to\eta_q$ are
from \cite{formfactorwangwei}, and  others from
\cite{formfactorStech}, all of which are shown in
Table~\ref{formfactor1}.

For the  final states involving $\eta$ or $\eta'$, it is convenient
to consider the flavor mixing of $\eta_q$ and $\eta_s$ with a mixing
angle $\phi$,
\begin{equation}\label{etamixing}
\left(
  \begin{array}{cc}
    \eta\\ \eta'
  \end{array}
\right) = \left(
  \begin{array}{cc}
    \cos\phi & -\sin\phi \\
    \sin\phi & \cos\phi
  \end{array}
\right) \left(
  \begin{array}{c}
    \eta_q \\ \eta_s
  \end{array}
\right),
\end{equation}
where $\eta_q$ and $\eta_s$ are defined by
\begin{eqnarray}
\eta_q=\frac{1}{\sqrt{2}}(u\bar{u}+d\bar{d}),&& \eta_s=s\bar{s}.
\end{eqnarray}
A recent experimental measurement from the KLOE collaboration gives
the mixing angle $\phi=(40.4\pm0.6)^{\circ}$\cite{KLOE}.

\begin{table}\caption{Meson decay constants (MeV). Those of $\pi$, $K$, $D$, and $D_s$ are
 from PDG\cite{PDG2010}, others are from
\cite{decayconstantsofVectors}.}\label{decayconstant}
\begin{tabular}{ccccccccccccc}
\hline\hline $f_{\pi}$ & $f_{K}$ & $f_{\rho}$ & $f_{K^*}$ &
$f_{\omega}$ & $f_{\phi}$ & $f_{D}$ & $f_{D_s}$
\\
\hline 130  & 156  & 216 & 220 & 187 & 215 & 207 & 258
\\\hline\hline
\end{tabular}
\end{table}

\begin{table}\caption{The $D$ meson transition form factors and dipole model parameters
$\alpha_{1,2}$.  The parameter $\alpha_1=\alpha_2+1$ if only
$\alpha_2$ is shown in the table. The form factors and $\alpha_i$
parameters  of $D$ to $\pi$ and $K$ are from \cite{formfactorCLEOc},
$D\to\eta_q$ from \cite{formfactorwangwei}, and  others from
\cite{formfactorStech}.}\label{formfactor1}
\begin{tabular}{ccccccccccccccc}
\hline\hline $ $ & $F_0^{D\pi}$ & $F_1^{D\pi}$ & $F_0^{DK}$ &
$F_1^{DK}$ & $F_1^{D_SK}$ & $F_1^{D_S\eta_s}$
  \\ \hline
$F(0)$ & 0.67 & 0.67 & 0.74 & 0.74 & 0.72  & 0.78
\\
$\alpha_2$ & 0.21 & 0.24 & 0.30 & 0.33 & 0.20 & 0.23
\\ \hline \hline
$ $  & $A_0^{D\rho}$& $A_0^{D\omega}$ & $A_0^{DK^*}$ &
$A_0^{D_SK^*}$ & $A_0^{D_S\phi}$
\\ \hline
$F(0)$  & 0.66 & 0.66 & 0.76 & 0.67 & 0.73
\\
$\alpha_2$  & 0.36 & 0.36 & 0.17 & 0.20 & 0.10
\\ \hline\hline
$ $ & $F_0^{D_S\to K}$& $F_0^{D_S\to\eta_s(M_{\eta})}$&
$F_0^{D_S\to\eta_s(M_{\eta'})}$ & $F_1^{D\to\eta_q}$  &
$F_0^{D\to\eta_q}$
\\ \hline
$F(0)$ & 0.72 & 0.78 & 0.78 & 0.69 & 0.69
\\
$\alpha_1$ & 0.41 & 0.33 & 0.21 & 1.03 & 0.39
\\
$\alpha_2$ & 0.70 & 0.38 & 0.76 & 0.29 & 0.01
\\ \hline\hline
\end{tabular}
\end{table}

The decay constants of $\eta$ or $\eta'$ are defined by
%\begin{equation}
%\langle 0|J^{i}_{\mu5}|P(p)\rangle=if^i_Pp_{\mu},
%\end{equation}
%where $i=q,s$ represent the flavor components, $P=\eta,\eta'$, and
%$J^i_{\mu5}$ denotes axial-vector current. So
\begin{eqnarray}
\langle
0|\bar{u}\gamma_{\mu}\gamma_5u|\eta(p)\rangle=if^u_{\eta}p_{\mu},&\langle
0|\bar{d}\gamma_{\mu}\gamma_5d|\eta(p)\rangle=if^d_{\eta}p_{\mu},&\langle
0|\bar{s}
\gamma_{\mu}\gamma_5s|\eta(p)\rangle=if^s_{\eta}p_{\mu},\nonumber\\
\langle
0|\bar{u}\gamma_{\mu}\gamma_5u|\eta'(p)\rangle=if^u_{\eta'}p_{\mu},&\langle
0|\bar{d}\gamma_{\mu}\gamma_5d|\eta'(p)\rangle=if^d_{\eta'}p_{\mu},&\langle
0|\bar{s} \gamma_{\mu}\gamma_5s|\eta'(p)\rangle=if^s_{\eta'}p_{\mu},
\end{eqnarray}
where
\begin{eqnarray}
f^u_{\eta}=f^d_{\eta}=\frac{1}{\sqrt{2}}f^q_{\eta},&&f^u_{\eta'}=f^d_{\eta'}=\frac{1}{\sqrt{2}}f^q_{\eta'}.
\end{eqnarray}
With the ansatz in \cite{kroll}, we have
\begin{eqnarray}
f^q_{\eta}=f_q\cos\phi,&& f^s_{\eta}=-f_s\sin\phi,
\nonumber\\
f^q_{\eta'}=f_q\sin\phi,&& f^s_{\eta'}=f_s\cos\phi .
\end{eqnarray}
%where,
%$<0|\bar{q}'\gamma_{\mu}\gamma_5q'|\eta_{q'}>=if_{q'}p_{\mu}$,
%$q'=q,s$.
It is assumed that $f_{q,s}$ obtained from the $\eta_{q,s}$
components of the wave functions are independent of the meson
involved. We use that $f_q=(1.07\pm0.02)f_{\pi}$ and
$f_s=(1.34\pm0.06)f_{\pi}$ from \cite{kroll}.

The form factors of $D\to\eta_q$ in Table \ref{formfactor1} denote
that of $D\to\eta_{u\bar{u},d\bar{d}}$, not
$D\to\frac{1}{\sqrt{2}}(u\bar{u}+d\bar{d})$,
%have the following relation
%\begin{eqnarray}
%F^{D^0\to\eta_{u\bar{u}}}=F^{D^+\to\eta_{d\bar{d}}}=F^{D\to\eta_q}.
%\end{eqnarray}
hence,
\begin{eqnarray}
F^{D\to\eta}=\frac{1}{\sqrt{2}} F^{D\to\eta_q}\cos\phi,&&
F^{D_S\to\eta}=-F^{D_S\to\eta_s}\sin\phi,\\
F^{D\to\eta'}=\frac{1}{\sqrt{2}}F^{D\to\eta_q}\sin\phi ,&&
F^{D_S\to\eta'}=F^{D_S\to\eta_s}\cos\phi .
\end{eqnarray}

In order to calculate the  annihilation-type diagrams in the pole
model, we have to know the effective strong coupling constants
between the intermediate state and two final states. Some of them
are obtained directly from experiments.
%For example,  we get $g_{\rho\pi\pi}=2.4$
%from the decay of $\rho^+\to\pi^+\pi^0$, and $g_{K^*K\pi}=2.6$ from
%$K^*(892)^0\to\pi^+K^-$.
Some others are related to the known ones using SU(3) symmetry.
Although the intermediate states are a little off shell,  in the
pole model they are used as on-shell resonant states. So these
on-shell strong couplings are used to calculate the annihilation
diagrams in an approximation.

There are many scalar mesons discovered by the experiments. The
existence of the lightest scalar nonet with the mass smaller than or
close to 1 GeV has been a problem for many years\cite{PDG2010}. It
is still controversial that they are primarily the four-quark bound
states or  two-quark scalar states.   In this work, we use
$K^*_0(1430)$, $a_0(1450)$, $f_0(1370)$,  and $f_0(1500)$
  as intermediate mesons in
the pole model for $D\to PP$ decays. The  decay constant of $K_0^*$
 is calculated in several methods, such as the finite-energy sum
 rule\cite{decayconstantsofscalar}, the generalized Nambu-Jona-Lasinio  model
\cite{decayconstantsofscalar2}, and so on. We shall use the results
in \cite{decayconstantsofscalar},
\begin{eqnarray}
 f_{K^*_0}=(42\pm2){\rm MeV}.
\end{eqnarray}
For all other scalar mesons,  we take the same value of decay
constants    as $K_0^*$ in the flavor SU(3) limit for
simplification.

The corresponding effective strong coupling constant between $K^*_0$
and the final states of $K\pi$ is evaluated by $g_{K^*_0K\pi}=2.7$
from the decay of $K^*_0(1430)^0\to\pi^-K^+$. Other couplings of
$g_{SPP}$ are of the same value in the SU(3) limit. For $D\to PV$
decays, the intermediate states are pseudoscalar mesons with
relatively large decay constants shown in Table \ref{decayconstant}.
 The corresponding effective strong
coupling constants are  $g_{\rho\pi\pi}=4.2$ obtained from
$\rho^+\to\pi^+\pi^0$ or $\rho^0\to\pi^+\pi^-$, $g_{K^*K\pi}=4.6$
from $K^*(892)^0\to\pi^+K^-$, and $g_{\phi KK}=4.5$ from $\phi\to
K^+K^-$. For decays involving $\eta$ or $\eta'$, we assume that
$g_q=4.2$, $g_s=4.6$, and $g_{ss}=4.5$, where $g_q$ couple to the
states with only $u$ or $d$ quarks, $g_s$ to two of the states with
$s$ quark, and $g_{ss}$ to all the three mesons with $s$ quark, so
that some effects of SU(3) breaking are considered.

\subsection{$D\to PP$}

For $D\to PP$ decays, the decay rate is
\begin{equation}
\Gamma(D\to PP)=\frac{p}{8\pi m_D^2}|\mathcal{A}|^2,
\end{equation}
where $p$ is the momentum of either meson in the final state in the
center-of-mass frame,
$p=\sqrt{(m_D^2-(m_{P_1}+m_{P_2})^2)(m_D^2-(m_{P_1}-m_{P_2})^2)}/{2m_D}$.

As is done in the naive factorization model, the Wilson coefficients
$a_i$'s are universal and process-independent, except with a
relative strong phase for different topological diagrams. Because of
the important nonperturbative effect of QCD in the charm system,
$a_1$ and $a_2$ should deviate a lot from the naive factorization
approach. In order to give the most suitable results, we input the
following values by hand,  which also used the hint from the fit of
diagrammatic approach:
\begin{equation} \label{eq:1}
 \begin{aligned}
         a_1 &= 1.25\pm0.10, \\
         a_2 &= (0.85\pm0.10) e^{i(153\pm10)^{\circ} }, \\
         a_A &= (0.90\pm0.10) e^{i(160\pm10)^{\circ} }, \\
         a_E &= (2.4\pm0.1) e^{i(55\pm10)^{\circ} }, \\
\end{aligned}
\end{equation}
where $a_1$ is the coefficient of the  color-favored emission tree
diagrams, $a_2$ for  the color-suppressed emission diagrams, $a_A$
for  the W-annihilation diagrams, and $a_E$ for  the W-exchange
diagrams.  Large relative strong phases are considered in the $a_i$,
due to unneglected inelastic final-state interactions in the $D$
decays.  These values of $a_1$ and $a_2$ are not far away from the
large $N_c$ limit, except that we use quite large strong phases,
which are required by the experimental data. As is discussed in the
diagrammatic approach \cite{diagrammaticchengchiang}, the
W-annihilation contributions with helicity-suppressed effect are
much smaller than those of the W-exchange diagrams. Therefore we use
a much larger coefficient of $a_E$ than  the W-annihilation
coefficient $a_A$.  Besides, we ignore the disconnected hairpin
diagrams, $SE$ and $SA$, as discussed in
\cite{diagrammaticchengchiang}.

\begin{table}\caption
{Branching ratios for Cabibbo-favored decays of $D\to PP$(\%). The
predicted branching ratios with both annihilation- and emission-type
contributions (all) and with only emission-type contributions
(emission) are given together with the  experimental
data\cite{dataofPP}, the recent results from the diagrammatic
approach \cite{diagrammaticchengchiang}, and the calculations
considering the final-state interaction (FSI) effects of nearby
resonances\cite{pietro} as comparison. }\label{CFPP}
\begin{tabular}{cccccccc}\hline\hline
 Modes & Br(FSI) & Br(diagrammatic) & Br(emission) & ~~~~~~Br(all)~~~~~~ &\
Br(exp)\\\hline
$D^+\to \pi^+\bar{K}^0$ & 2.51 & 3.08$\pm $0.36 & $3.1\pm2.0$ & $3.1\pm2.0$ & 3.074$\pm $0.096\\
$D^0\to \pi^+K^-$       & 4.03 & 3.91$\pm $0.17 & $5.8\pm0.7$ & $3.9\pm1.0$ & 3.891$\pm $0.077\\
$D^0\to \pi^0\bar{K}^0$ & 1.35 & 2.36$\pm $0.08 & $2.1\pm0.6$ & $2.4\pm0.7$ & 2.38$\pm $0.09\\
$D^0\to \bar{K}^0\eta$  & 0.80 & 0.98$\pm $0.05 & $0.9\pm0.2$ & $0.8\pm0.2$ & 0.96$\pm $0.06\\
$D^0\to \bar{K}^0\eta'$ & 1.51 & 1.91$\pm $0.09 & $0.3\pm0.2$ & $1.9\pm0.3$ & 1.90$\pm $0.11\\
$D_S^+\to K^+\bar{K}^0$ & 4.79 & 2.97$\pm $0.32 & $5.1\pm0.9$ & $3.0\pm0.9$ & 2.98$\pm $0.08\\
$D_S^+\to \pi^+\eta$    & 1.33 & 1.82$\pm $0.32 & $3.8\pm0.4$ & $1.9\pm0.5$ & 1.84$\pm $0.15\\
$D_S^+\to \pi^+\eta'$   & 5.89 & 3.82$\pm $0.36 & $2.9\pm0.6$ & $4.6\pm0.6$ & 3.95$\pm $0.34\\
$D_S^+\to \pi^+\pi^0$& & 0&0&0&$<$0.06\\
\hline\hline\end{tabular} \end{table}

The predicted branching ratios with annihilation-type contributions
(all) and without annihilation type contributions (emission)
together with experimental measurements of charmed mesons decay into
two pseudoscalar mesons are listed in Tables~\ref{CFPP},
\ref{SCSPP}, \ref{DCSPP}, for  the Cabibbo-favored decays,  the
singly Cabibbo-suppressed decays, and  the doubly Cabibbo-suppressed
decays, respectively. There are many sources of theoretical
uncertainties in the calculations.  Since the decay constants of
pseudoscalar and vector mesons are taken from experiments with very
small errors, our numerical results are not very sensitive to the
variations of meson decay constants. The branching ratios are truly
sensitive to the coefficients of $a_1$ and $a_2$, especially to
their relative strong phases.
 Since the systematic errors from
theoretical models are usually difficult to estimate, we show
uncertainties
 at the tables only from the parameters $a_i$ in
Eq.(\ref{eq:1}). For comparison we also show the recent results from
the diagrammatic approach \cite{diagrammaticchengchiang} and those
considering final-state interaction effects of nearby resonances
\cite{pietro}. It is clear that our results with large
annihilation-type contributions agree with experiments much better
than that of Ref.\cite{pietro}. For the Cabibbo-favored channels,
which are the input data for $\chi^2$ fit in the diagrammatic
approach, we have comparable results with the diagrammatic approach
\cite{diagrammaticchengchiang}. For other channels, we have better
agreement with experiments than the diagrammatic approach. The
reason is mostly due to the SU(3) breaking effects, which had been
fully neglected in the diagrammatic approach.

\begin{table}\caption
{Same as Table.\ref{CFPP} except for singly Cabibbo-suppressed
decays of $D\to PP$($\times10^{-3}$).}\label{SCSPP}
\begin{tabular}{cccccccc}\hline\hline
 Modes & Br(FSI) & Br(diagrammatic) & Br(emission) & ~~~~~~Br(all)~~~~~~  &
Br(exp)\\\hline
$D^+\to \pi^+\pi^0$  & 1.7 & 0.88$\pm $0.10 & $1.0\pm0.5$ & $1.0\pm0.5$ & 1.18$\pm $0.07\\
$D^+\to K^+\bar{K}^0$& 8.6 & 5.46$\pm $0.53 & $11.3\pm1.6$ & $8.4\pm1.6$ & 6.12$\pm $0.22\\
$D^+\to \pi^+\eta$   & 3.6 & 1.48$\pm $0.26 & $3.1\pm1.0$ & $1.6\pm1.0$ & 3.54$\pm $0.21\\
$D^+\to \pi^+\eta'$  & 7.9 & 3.70$\pm $0.37 & $3.7\pm0.7$ & $5.5\pm0.8$ & 4.68$\pm $0.29\\
$D^0\to \pi^+\pi^-$  & 1.59 & 2.24$\pm $0.10 & $3.0\pm0.4$ & $2.2\pm0.5$ & 1.45$\pm $0.05\\
$D^0\to \pi^0\pi^0$  & 1.16 & 1.35$\pm $0.05 & $0.7\pm0.2$ & $0.8\pm0.2$ & 0.81$\pm $0.05\\
$D^0\to K^+K^-$      & 4.56 & 1.92$\pm $0.08 & $4.4\pm0.5$ & $3.0\pm0.8$ & 4.07$\pm $0.10\\
$D^0\to K^0\bar{K}^0$& 0.93 & 0&0& $0.3\pm0.1$ &0.64$\pm $0.08\\
$D^0\to \pi^0\eta$   & 0.58 & 0.75$\pm $0.02 & $0.7\pm0.2$ & $1.1\pm0.3$ & 0.68$\pm $0.07\\
$D^0\to \pi^0\eta'$  & 1.7 & 0.74$\pm $0.02 & $0.6\pm0.1$ & $0.6\pm0.2$ & 0.91$\pm $0.13\\
$D^0\to \eta\eta$    & 1.0 & 1.44$\pm $0.08 & $1.3\pm0.4$ & $1.3\pm0.4$ & 1.67$\pm $0.18\\
$D^0\to \eta\eta'$   & 2.2 & 1.19$\pm $0.07 &$0.04\pm0.04$&$1.1\pm0.1$&1.05$\pm $0.26\\
$D_S^+\to \pi^0K^+$  & 1.6 & 0.86$\pm $0.09 & $0.9\pm0.2$ & $0.5\pm0.2$ & 0.62$\pm $0.23\\
$D_S^+\to \pi^+K^0$  & 4.3 & 2.73$\pm $0.26 & $4.1\pm0.5$ & $2.8\pm0.6$ & 2.52$\pm $0.27\\
$D_S^+\to K^+\eta$   & 2.7 & 0.78$\pm $0.09 & $0.8\pm0.5$ & $0.8\pm0.5$ & 1.76$\pm $0.36\\
$D_S^+\to K^+\eta'$  & 5.2 & 1.07$\pm $0.17 & $0.7\pm0.3$ & $1.4\pm0.4$ & 1.8$\pm $0.5\\
\hline\hline\end{tabular} \end{table}

\begin{table}\caption
{Same as Table.\ref{CFPP} except for doubly Cabibbo-suppressed
decays of $D\to PP$($\times10^{-4}$). }\label{DCSPP}
\begin{tabular}{cccccccc}\hline\hline
Modes & Br(diagrammatic) & Br(emission)  &~~~~~~Br(all)~~~~~~ &
Br(exp)\\\hline
$D^+\to \pi^+K^0$ & 1.98$\pm $0.22 & $2.8\pm0.5$ & $1.7\pm0.5$ &  \\
$D^+\to \pi^0K^+$ & 1.59$\pm $0.15 & $3.0\pm0.4$ & $2.2\pm0.4$ & 1.72$\pm $0.19\\
$D^+\to K^+\eta$  & 0.98$\pm $0.04 & $1.3\pm0.2$ & $1.2\pm0.2$ & 1.08$\pm$0.17\footnote{Data from \cite{won}}\\
$D^+\to K^+\eta'$ & 0.91$\pm $0.17 & $0.4\pm0.1$ & $1.0\pm0.1$ & 1.76$\pm$0.22\footnote{Data from \cite{won}}\\
$D^0\to \pi^0K^0$ & 0.67$\pm $0.02 & $0.5\pm0.2$ & $0.6\pm0.2$ &  \\
$D^0\to \pi^-K^+$ & 1.12$\pm $0.05 & $2.3\pm0.3$ & $1.6\pm0.4$ & 1.48$\pm $0.07\\
$D^0\to K^0\eta$  & 0.28$\pm $0.02 &$0.23\pm0.05$&$0.22\pm0.05$& \\
$D^0\to K^0\eta'$ & 0.55$\pm $0.03 &$0.08\pm0.06$&$0.5\pm0.1$& \\
$D_S^+\to K^+K^0$ & 0.38$\pm $0.04 & $0.7\pm0.4$ & $0.7\pm0.4$ &   \\
\hline\hline\end{tabular} \end{table}

The branching ratio of the   pure annihilation process $D_s^+\to
\pi^+\pi^0$  is vanished in our pole model. The resonant state that
annihilates to $\pi^+\pi^0$ is a scalar meson ($0^{++}$), whose
isospin could be 0, 1, or 2. However, isospin-0 would be ruled out
because of charged final states, and isospin-2 is forbidden for the
leading order $\Delta C=1$ weak decay. For the case of isospin-1,
its $G$ parity would be odd, which conflicts to a system of two
pions whose $G$ parity is even. Therefore, no resonant states can be
produced and then annihilate to $\pi^+\pi^0$. In another word, no
annihilation diagrams contribute to $D_s^+\to \pi^+\pi^0$ and
$D^+\to \pi^+\pi^0$. In fact, this kind of contribution is forbidden
from the isospin symmetry of $\pi^+$ and $\pi^0$ as identical
particles. Simply, two pions can not form an $s$-wave isospin 1
state, because of the Bose-Einstein statics.

{ The pure annihilation process $D^0\to K^0\bar K^0$, with nonzero
experimental data, also demonstrates the important annihilation-type
contributions. There are two kinds of contributions to this mode
with  $d\bar d$ and $s\bar s$  produced from weak vertex,
respectively. In the flavor SU(3) limit, the rate vanishes due to
the cancelation of CKM matrix elements, as predicted in the
diagrammatic approach. Therefore, the effect of the SU(3) breaking
is the dominant contribution here. In our pole model, we use
$f_0(1370)$ and $f_0(1500)$ as two different poles with the $d\bar
d$ and $s\bar s$ components, respectively, to describe the
corresponding SU(3) breaking effect. We also refer to the argument
of the long distance resonance effect in
\cite{diagrammaticchengchiang,cheng}, the $t$-channel final-state
interaction in \cite{D0K0K0bartchannelFSI}, the nonfactorizable
chiral loop contributions in \cite{D0K0K0barchiralloop}, and the
SU(3) breaking effect in the effective Wilson coefficients in
\cite{a1a2phaseyufengzhou} for this channel.

Large branching ratios with $\eta'$ in the final states are both
measured  and predicted, which are larger than those with $\eta$ in
most cases, such as $D^0$ decays into $\bar{K}^0\eta$,
$\bar{K}^0\eta'$ and $D_s^+$ into $\pi^+\eta$, $\pi^+\eta'$,
although the phase spaces with $\eta'$ are smaller than those with
$\eta$. For $\eta$, the contributions from the components of $d\bar
d$ and $s\bar s$ are destructive due to the minus sign in the mixing
matrix of Eq.(\ref{etamixing}) and the positive mixing
angle\footnote{The theoretical and phenomenological estimates for
mixing angle $\phi$ is $42.2^\circ$ and $(39.3 \pm 1.0)^\circ$,
respectively \cite{kroll}}; while they are constructive for $\eta'$.
Besides, large $W$-exchange contributions dominate most $\eta'$
modes, especially for $D^0\to\eta\eta'$ and $D^0\to\bar K^0\eta'$,
which demonstrates large annihilation-type contributions directly
again. We also refer to this issue with $K_0^*(1430)$ as an
resonance in the spacelike form factors in the factorization
approach in \cite{fajferK1340}, some effects of the inelastic
final-state interactions in \cite{etafsi}, the final-state phases of
the amplitudes in \cite{Rosner:1999xd}, and the two-gluon anomaly
effects in \cite{etaanomaly}.
 }

\subsection{$D\to PV$ decays}

The decay rate of $D\to PV$ decays is
\begin{equation}
\Gamma(D\to PV)=\frac{p}{8\pi m_D^2}\sum_{pol.}|\mathcal{A}|^2,
\end{equation}
by summing over all the polarization states of the vector mesons.

We assume that the coefficients $a_i$ are universal and process
independent for $D\to PV$, but they are different from those of
$D\to PP$ as discussed in \cite{a1a2phasehaiyangcheng} and
\cite{diagrammaticchengchiang}. Their absolute values are larger
than those of $D\to PP$ because the soft final-state interactions
make more effects on $D\to PV$ decays. In our calculations, they are
used as
\begin{equation} \label{eq:2}
\begin{aligned}
         a_1^{PV} &= 1.32\pm0.10, \\
         a_2^{PV} &= (0.75\pm0.10) e^{i(160\pm10)^{\circ} }, \\
         a_A^{PV} &= (0.12\pm0.10) e^{i(345\pm10)^{\circ} }, \\
         a_E^{PV} &= (0.62\pm0.10) e^{i(238\pm10)^{\circ} }.
\end{aligned}
\end{equation}
Similarly to the $PP$ modes, large relative strong phases due to
inelastic final-state interactions are considered in the $a_i$.
Again, the contributions from W-annihilation diagrams are smaller
than the W-exchange ones. Besides, the relative strong phase between
$a_1^{PV}$ and $a_2^{PV}$ is in accordance with the results from the
diagrammatic
approach\cite{diagrammaticchengchiang,a1a2phasehaiyangcheng}.

\begin{table}\caption
{Branching ratios for Cabibbo-favored decays of $D\to PV$(\%). The
predicted rates with only emission-type contributions (emission) and
with both annihilation- and emission-type contributions (all) are
shown in the table, compared with the  experimental
data\cite{PDG2010}, the  fitted results from the  diagrammatic
approach\cite{diagrammaticchengchiang} in which only the (A, A1)
solution is quoted, and the results considering final-state
interaction (FSI) effects of nearby resonances\cite{pietro}.}
\label{CFPV}\begin{tabular}{cccccccc}\hline\hline Modes & Br(FSI) &
Br(diagrammatic) & Br(emission) &~~~~~~Br(all)~~~~~~ &\
Br(exp)\\\hline
$D^0\to K^-\rho^+$         & 11.19 & 10.8$\pm $2.2 & $12.2\pm1.8$ & $8.8\pm2.2$  &  10.8$\pm $0.7\\
$D^0\to \bar{K}^0\rho^0$   & 0.88 & 1.54$\pm $1.15& $0.7\pm0.5$ & $1.7\pm  0.7$ & $1.32^{+0.12}_{-0.16}$\\
$D^0\to \pi^0\bar{K}^{*0}$ & 3.49 & 2.82$\pm $0.34& $2.3\pm0.7$ & $2.9\pm  1.0$ & 2.82$\pm $0.35\\
$D^0\to \pi^+{K}^{*-}$     & 4.69 & 5.91$\pm$0.70 & $3.8\pm0.7$ & $3.1\pm  1.0$ & $5.68^{+0.68}_{-0.53}$\\
$D^0\to \eta\bar{K}^{*0}$  & 0.51 & 0.96$\pm $0.32& $0.7\pm0.2$ & $0.7\pm  0.2$ & 0.96$\pm $0.30\\
$D^0\to \eta'\bar{K}^{*0}$ & 0.005 & 0.012$\pm $0.003&$0.003\pm0.001$&$  0.016\pm0.005$&$< $0.11\\
$D^0\to \bar{K}^0\omega$   & 2.16 & 2.26$\pm $1.38& $0.6\pm0.5$ & $2.5\pm  0.7$ & 2.22$\pm $0.12\\
$D^0\to \bar{K}^0\phi$     & 0.90 & 0.868$\pm $0.139& $0$        & $0.8\pm0.2$   &  0.868$\pm $0.060\\
$D^+\to \pi^+\bar{K}^{*0}$ & 0.64 & 1.83$\pm $0.49& $1.4\pm1.3$ & $1.4\pm  1.3$ & 1.56$\pm $0.18\\
$D^+\to \bar{K}^0\rho^+$   & 11.77 & 9.2$\pm $6.7 & $15.1\pm3.8$  & $15.1\pm3.8  $ &9.4$\pm $2.0\\
$D_S^+\to K^+\bar{K}^{*0}$ & 3.86 &             & $5.6\pm1.9$ & $4.2\pm1.7$   &  3.90$\pm $0.23\\
$D_S^+\to \bar{K}^0 {K}^{*+}$&3.37 &           & $1.7\pm0.7$ & $1.0\pm0.6$   &  5.4$\pm $1.2\\
$D_S^+\to \eta\rho^+$&     9.49    &           & $8.3\pm1.3$ & $8.3\pm1.3$   &  8.9$\pm $0.8\cite{dataofDstorhoeta}\\
$D_S^+\to \eta'\rho^+$&      2.61  &           & $3.0\pm0.5$ & $3.0\pm0.5$   &  12.2$\pm $2.0\\
$D_S^+\to \pi^+\phi$      & 2.89  & 4.38$\pm $0.35 & $4.3\pm0.6$ & $4.3\pm 0.6$ & 4.5$\pm $0.4\\
$D_S^+\to \pi^+\rho^0$&       0.080  &          & $0$&$0.4\pm0.4$&0.02$\pm $0.012\\
$D_S^+\to \pi^0\rho^+$     & 0.080 & &$0$&$0.4\pm0.4$& \\
$D_S^+\to \pi^+\omega$     & 0.0 & &$0$&$0$&0.23$\pm $0.06\\
\hline\hline\end{tabular} \end{table}

\begin{table}\caption
{Same as Table.\ref{CFPV} except for singly Cabibbo-suppressed
decays of $D\to PV$($\times\ 10^{-3}$)}
\label{SCSPV}\begin{tabular}{cccccccc}\hline\hline  Modes & Br(FSI)
& Br(diagrammatic) &  Br(emission)  &~~~~~~Br(all)~~~~~~ &
Br(exp)\\\hline
$D^0\to \pi^-\rho^+$ & 8.2 & 8.34$\pm $1.69 & $7.4\pm1.3$ & $10.2\pm1.5$   &  9.8$\pm $0.4\\
$D^0\to \pi^+\rho^-$ & 6.5 & 3.92$\pm $0.46 & $1.8\pm0.5$ & $3.5\pm0.6  $ & 4.97$\pm $0.23\\
$D^0\to \pi^0\rho^0$ & 1.7 &  2.96$\pm $0.98 & $1.4\pm0.6$ & $1.4\pm 0.6$  & 3.73$\pm $0.22\\
$D^0\to K^- {K}^{*+}$& 4.5 &  4.25$\pm $0.86 & $5.5\pm0.8$ & $4.7\pm0.8  $ & 4.38$\pm $0.21\\
$D^0\to K^+ {K}^{*-}$& 2.8 &  1.99$\pm $0.24 & $2.0\pm0.3$ & $1.6\pm  0.3$ & 1.56$\pm $0.12\\
$D^0\to \bar{K}^0K^{*0}$ & 0.99 &  0.29$\pm $0.22 & 0 & $0.16\pm0.05$ &  $<$0.9\\
$D^0\to K^0\bar{K}^{*0}$ & 0.99 &  0.29$\pm $0.22 & 0 &$ 0.16\pm0.05$ &  $<$1.8\\
$D^0\to \pi^0\omega$ & 0.08 &  0.10$\pm $0.18 & $0.08\pm0.02$ & $0.08\pm0.02$ & $<$0.26\\
$D^0\to \pi^0\phi$   & 1.1 &  1.22$\pm $0.08 & $1.0\pm0.3$   & $1.0\pm0.3$   & 0.76$\pm $0.05\\
$D^0\to \eta\phi$    & 0.57 &  0.31$\pm $0.10 & $0.23\pm0.06$   & $0.23\pm0.06$   &  0.14$\pm $0.05\\
$D^0\to \eta\rho^0$  & 0.24 &  1.11$\pm $0.86 & $0.05\pm0.01$ & $0.05\pm0.01$ & \\
$D^0\to \eta'\rho^0$ & 0.10 &  0.14$\pm $0.02 & $0.08\pm0.02$ & $0.08\pm0.02$ & \\
$D^0\to \eta\omega$  & 1.9 &  3.08$\pm$1.42  & $1.2\pm0.3$   & $1.2\pm0.3 $  & 2.21$\pm $0.23 \\
$D^0\to \eta'\omega$ & 0.001 &  0.07$\pm$0.02  & $0.0001\pm0.0001$ & $0.0001 \pm0.0001$ &  \\
$D^+\to \pi^+\rho^0$ & 1.7 &  &$0.4\pm0.4$ & $0.8\pm0.7$ &  0.83$\pm $0.15\\
$D^+\to \pi^0\rho^+$ & 3.7 &  &$5.3\pm1.7$ & $3.5\pm1.6$ & \\
$D^+\to K^+\bar{K}^{*0}$   & 2.5 &  & $5.1\pm1.1$ & $4.1\pm1.0$ & $3.76^{+0.20}_{-0.26}$\\
$D^+\to \bar{K}^0 {K}^{*+}$& 1.70 &  & $14.0\pm2.5$ & $12.4\pm2.4$ &  32$\pm $14\\
$D^+\to \eta\rho^+$        & 0.002 &  & $0.4\pm0.4$ & $0.4\pm0.4$ &   $< $ 7\\
$D^+\to \eta'\rho^+$       & 1.3 &  & $0.8\pm0.1$   & $0.8\pm0.1$   &  $<$5\\
$D^+\to \pi^+\phi$         & 5.9 & 6.21$\pm $0.43 & $5.1\pm1.4$ & $5.1\pm1.4$ &  5.44$\pm $0.26\\
$D^+\to \pi^+\omega$       & 0.35 &  & $0.3\pm0.3$   & $0.3\pm0.3$   &  $<$0.34\\
$D_S^+\to \pi^+K^{*0}$     & 3.3 &  & $2.3\pm0.8$ & $1.5\pm0.7$ &  2.25$\pm $0.39\\
$D_S^+\to \pi^0 {K}^{*+}$  & 0.29 &  & $0.4\pm0.2$ & $0.1\pm0.1$ & \\
$D_S^+\to K^+\rho^0$       & 2.4 &  & $1.6\pm0.6$ & $1.0\pm0.6$ & 2.7$\pm $0.5  \\
$D_S^+\to K^0\rho^+$       & 19.5 &  & $9.7\pm2.2$ & $7.5\pm2.1$ & \\
$D_S^+\to \eta {K}^{*+}$   & 0.24 &  & $1.0\pm0.4$ & $1.0\pm0.4$ & \\
$D_S^+\to \eta' {K}^{*+}$  & 0.24 &  & $0.4\pm0.2$ & $0.6\pm0.2$ & \\
$D_S^+\to K^+\omega$       & 0.72 &  & $1.1\pm0.7$ & $1.8\pm0.7$ & $ <$2.4 \\
$D_S^+\to K^+\phi$         & 0.15 &  & $0.3\pm0.3$ & $0.3\pm0.3$ & $<$0.6\\
\hline\hline\end{tabular} \end{table}

\begin{table}\caption
{Same as Table.\ref{CFPV} except for doubly Cabibbo-suppressed
decays of $D\to PV$($\times 10^{-4}$)}
\label{DCSPV}\begin{tabular}{cccccccc}\hline\hline  Modes &
Br(diagrammatic) &  Br(emission)  &~~~~~~Br(all)~~~~~~ &
Br(exp)\\\hline
$D^0\to \pi^- {K}^{*+}$ & 3.59$\pm $0.72 & $3.7\pm0.6$ & $2.7\pm0.6$ & $3.54^{+1.80}_{-1.05}$\\
$D^0\to \pi^0K^{*0}$    & 0.54$\pm $0.18 & $0.6\pm0.2$ & $0.8\pm0.3$ & \\
$D^0\to K^+\rho^-$      & 1.45$\pm $0.17 & $1.1\pm0.2$ & $0.9\pm0.3$ & \\
$D^0\to K^0\rho^0$      & 0.91$\pm $0.51 & $0.2\pm0.1$ & $0.5\pm0.2$ & \\
$D^0\to K^0\omega$      & 0.58$\pm $0.40 & $0.2\pm0.1$ & $0.7\pm0.2$ & \\
$D^0\to K^0\phi  $      & 0.06$\pm $0.05 & 0           & $0.20\pm0.06$ & \\
$D^0\to \eta K^{*0}$    & 0.33$\pm $0.08 & $0.18\pm0.05$ & $0.17\pm0.05$ & \\
$D^0\to \eta' K^{*0}$      & 0.0040$\pm $0.0006 & $0.001\pm0.001$ & $0.004\pm0.001$ & \\
$D^+\to \pi^+K^{*0}$ &  & $3.0\pm1.0$ & $2.2\pm0.9$ &  3.75$\pm $0.75\\
$D^+\to \pi^0K^{*+}$ &  & $4.7\pm0.9$ & $4.0\pm0.9$ &  \\
$D^+\to K^+\rho^0$   &  & $1.4\pm0.4$&$1.0\pm0.4$&2.1$\pm $0.5  \\
$D^+\to K^0\rho^+$   &  & $0.9\pm0.4$ & $0.5\pm0.4$ &  \\
$D^+\to K^+\omega$   &  & $1.4\pm0.5$ & $1.8\pm0.5$ &  \\
$D^+\to K^+\phi  $   &  & 0 & $0.2\pm0.2$ &  \\
$D^+\to \eta K^{*+}$ &  & $1.5\pm0.2$ & $1.4\pm0.2$ &  \\
$D^+\to \eta' K^{*+}$ &  & $0.013\pm0.006$ & $0.020\pm0.07$ &  \\
$D_S^+\to K^+K^{*0}$&0.20$\pm $0.05&$0.2\pm0.2$&$0.2\pm  0.2$& \\
$D_S^+\to K^0 {K}^{*+}$&1.17$\pm $0.86&$2.3\pm0.6$&$2.3\pm  0.6$& \\
\hline\hline\end{tabular} \end{table}

Our prediction of branching ratios of  the Cabibbo-favored,  the
singly Cabibbo-suppressed and  the doubly Cabibbo-suppressed  $D\to
PV$ decays are shown in Tables~\ref{CFPV}, \ref{SCSPV}, and
\ref{DCSPV}, respectively. The results in the third column
(emission) in each of these tables are the predictions of rates with
only the emission-type processes; while the results in the fourth
column (all) also include the annihilation-type contributions. It is
obvious that the annihilation-type contributions are of the same
order as  the emission-type diagrams, since the intermediate states
here in the pole model are pseudoscalar mesons with relatively
larger decay constants than those scalar mesons of the $D\to PP$
case. Again, for theoretical uncertainty estimation, we use only
those from the parameters  $a_i$ shown in Eq.(\ref{eq:2}), as
illustration. For comparison, we also list the results of the
diagrammatic approach \cite{diagrammaticchengchiang} and the
experimental date in these tables. It is easy to see that our
results with the annihilation-type contributions agree with the
experimental data. This means that the single-pole contribution
dominates the annihilation-type contribution in most $D\to PV$ decay
channels. For example, although the $D^0 \to\bar K^0 \phi$ channel
has no emission-type contribution, with vanishing branching ratio in
the factorization approach, our pole model gives the right branching
ratios agreeing with the experiment. This also confirms the
calculation done in the perturbative QCD approach \cite{du}.
Besides, some of the SU(3) flavor symmetry breaking effects are
considered in this work since the decay constants, transition form
factors and effective strong coupling constants are involved.

There is no resonant state contributing to the W-exchange diagram of
$D^0\to\pi^0\rho^0$ in the pole model, because a $\pi^0$ would
violate the $C$ parity, similarly to the case of
$D^0\to\eta(\eta')\omega,\eta\phi$. Besides, the single-pole
annihilation diagrams can not contribute to the
$D\to\rho\eta,\pi\omega$ decays because of $G$ parity violation. The
isospin of resonant state for $\rho\eta$ or $\pi\omega$ is one, so
the $G$ parity of the intermediate state is odd since it is a
pseudoscalar meson. However, the $G$ parity of $\rho$ and $\eta$ are
both even, and that of $\pi$ and $\omega$ are both odd, so the total
$G$ parity of  the final states is even. Therefore, no resonant
states are available for the  decays of $D$ mesons into $\rho\eta$
and $\pi\omega$. It is even worse for the pure annihilation process
$D^+_s\to\pi^+\omega$, since its decay rate is predicted to be zero
in the single-pole model, but it is not small in  the experiment.
This has already been discussed that this channel may be dominated
by the final-state rescattering via quark exchange in
\cite{diagrammaticchengchiang,cheng}, and by hidden strangeness
final-state interactions in \cite{fajfer}. Besides, the pure
annihilation mode $D_s^+\to\pi^+\rho^0$ is predicted much larger in
the pole model than the experiment data. The contributions from the
two diagrams in this channel are constructive since the minus sign
in the normalization of $\rho^0$ is compensated by the asymmetric
space wave function of the two final states which are in the
$P$-wave state. These two channels make such trouble that we fail to
find a reasonable solution of $A_P$ and $A_V$ and predict the $PV$
modes with the W-annihilation contributions in the diagrammatic
approach\cite{diagrammaticchengchiang}. Hence, further discussions
are still needed for these two channels.

$D^+_S\to \rho\eta'$ is predicted much smaller than the mode of
$D^+_S\to \rho\eta$, but the experimental branching ratios of the
former is larger. This is a puzzle that the  phase space of the
former mode is much smaller than the latter, so its branching ratio
should be smaller. In fact, the experimental measurement of
$D^+_S\to \rho\eta'$ \cite{cleo} is already too old. It is already
questioned by the PDG \cite{PDG2010}, since this branching fraction
$(12.5\pm2.2)\%$ considerably exceeds the recent inclusive $\eta'$
fraction of $(11.7\pm1.8)\%$.

\section{Summary }

We have calculated the branching ratios for the two-body hadronic
decays of charmed mesons into $PP$ and $PV$ using the generalized
factorization approach for the  emission-type diagrams and the
pole-dominance model for  the annihilation-type diagrams. Relative
strong phases between different topological diagrams, which are
important in the charmed decays, are considered in this work. Most
of our predicted branching fractions are  in accordance with  the
experimental data. Besides, compared to the naive and generalized
factorization models ever before, the results in this work are much
better since we have considered  the annihilation-type diagrams and
the relative strong phases between diagrams.

 We find that the annihilation-type contributions in the
pole model are large for both $PP$ and $PV$ modes, which is also
indicated by the difference between the life time of $D^+$ and
$D^0$. Comparing with the model-independent diagrammatic approach,
we reproduce their results with our specific model considering some
SU(3) breaking effects. Furthermore, we get more predictions in many
$D\to PV$ decay channels, which are absent in the diagrammatic
approach\cite{diagrammaticchengchiang}. Most of the results have a
better agreement with experimental data than previous calculations.

\section*{Acknowledgment}

We are grateful to Hai-Yang Cheng, Wei Wang, Yu-Ming Wang,
Dong-Sheng Du, Ping Wang, Qiang Zhao, Run-Hui Li, and Cheng Li for
useful discussions. This work is partially supported by National
Natural Science Foundation of China under the Grants No. 10735080
 and No. 11075168; National Basic Research Program of China (973)
  No. 2010CB833000; Natural Science Foundation of Zhejiang Province of China,
  Grant No. Y606252, and Scientific Research Fund of Zhejiang Provincial Education Department of China, Grant No. 20051357.

%%%%%%%%%%%%%%%%%%%%%%%%%%%%%%%%%%%%%%%%%%%%%%%%%%%%%%%%%%%%%%%%%%%%%%%%%%%%%%%%%%%%%%%%%%%%%%%%%%%%%%%%%%%%%%%%%%%%%%%%%%%%%%%%%%%%%%%%%%%%%%%%%%%%%%%%%%%%
\appendix
\section{Individual formulas for various decay channels of $D$ mesons}

The different contribution formulas for Cabibbo-favored decays of
$D\to PP$  are listed as
\begin{eqnarray}
\mathcal{A}(D^+\to\pi^+\bar{K}^0)&=&i\frac{G_F}{\sqrt{2}}V_{cs}^*V_{ud}
\bigg( a_2 f_{K} (m_D^2-m_{\pi}^2) F_0^{D\pi}(m_K^2)+ a_1 f_{\pi}
(m_D^2-m_K^2) F_0^{DK}(m_{\pi}^2)\bigg),
\nonumber\\
\mathcal{A}(D^0\to\pi^+K^-)&=&i\frac{G_F}{\sqrt{2}}V_{cs}^*V_{ud}
\bigg( a_1 f_{\pi} (m_D^2-m_K^2) F_0^{DK}(m_{\pi}^2)-a_E g_{1} f_S
f_D \frac{m_D^2m_{K^*_0}}{m_D^2-m_{K^*_0}^2}\bigg),
\nonumber\\
\mathcal{A}(D^0\to \pi^0\bar{K}^0)&=& i\frac{G_F}{2}V_{cs}^*V_{ud}
\bigg( a_E g_{1}f_S f_D \frac{m_D^2m_{K^*_0}}{m_D^2-m_{K^*_0}^2} +
a_2 f_{K} (m_D^2-m_{\pi}^2) F_0^{D\pi}(m_{K}^2)\bigg),
\nonumber\\
\mathcal{A}(D^0\to \bar{K}^0\eta)&=& i\frac{G_F}{2}V_{cs}^*V_{ud}
\bigg(a_2 f_{K} (m_D^2-m_{\eta}^2) F_0^{D\eta_q}(m_{K}^2)\cos\phi-
a_E g_{1}f_S f_D \frac{m_D^2m_{K^*_0}}{m_D^2-m_{K^*_0}^2}
(\cos\phi-\sqrt{2}\sin\phi)\bigg),
\nonumber\\
\mathcal{A}(D^0\to \bar{K}^0\eta')&=& i\frac{G_F}{2}V_{cs}^*V_{ud}
\bigg(a_2 f_{K} (m_D^2-m_{\eta'}^2) F_0^{D\eta_q}(m_{K}^2)\sin\phi-
a_E g_{1} f_S f_D\frac{m_D^2m_{K^*_0}}{m_D^2-m_{K^*_0}^2}
(\sin\phi+\sqrt{2}\cos\phi) \bigg),
\nonumber\\
\mathcal{A}(D_S^+\to K^+\bar{K}^0)&=&
i\frac{G_F}{\sqrt{2}}V_{cs}^*V_{ud} \bigg( a_2 f_{K}
(m_{D_S}^2-m_{K}^2) F_0^{D_SK}(m_{K}^2)- a_A g_{1}f_S f_{D_S}
\frac{m_{D_S}^2m_{a_0}}{m_{D_S}^2-m_{a_0}^2} \bigg),
\nonumber\\
\mathcal{A}(D_S^+\to \pi^+\eta)&=& -iG_FV_{cs}^*V_{ud} \bigg(
\frac{1}{\sqrt{2}}a_1 f_{\pi} (m_{D_S}^2-m_{\eta}^2)
F_0^{{D_S}\eta_s}(m_{\pi}^2)\sin\phi +a_A g_{1} f_S f_{D_S}
\frac{m_{D_S}^2m_{a_0}}{m_{D_S}^2-m_{a_0}^2}\cos\phi\bigg),
\nonumber\\
\mathcal{A}(D_S^+\to \pi^+\eta')&=& iG_FV_{cs}^*V_{ud}
\bigg(\frac{1}{\sqrt{2}}a_1 f_{\pi} (m_{D_S}^2-m_{\eta'}^2)
F_0^{{D_S}\eta_s}(m_{\pi}^2)\cos\phi- a_A g_{1} f_S f_{D_S}
\frac{m_{D_S}^2m_{a_0}}{m_{D_S}^2-m_{a_0}^2}\sin\phi\bigg),
\end{eqnarray}
where $f_S$ and $g_1=2.7$ are, respectively, denoted as the decay
constant of scalar mesons and effective strong coupling constant
between the intermediate state and final states in the limit of
SU(3) symmetry. Some phases from the propagators of the
 intermediate resonances are absorbed in the effective Wilson coefficients
 $a_A$
and $a_E$.

 The formulas for singly Cabibbo-suppressed
decays of $D\to PP$ are shown as {
\begin{eqnarray}
\mathcal{A}(D^+\to \pi^+\pi^0)&=& -i\frac{G_F}{2}V_{cd}^*V_{ud}
f_{\pi}(m_D^2-m_{\pi}^2)F_0^{D\pi}(m_{\pi}^2)(a_1+a_2),
\nonumber\\
\mathcal{A}(D^+\to K^+\bar{K}^0)&=&
i\frac{G_F}{\sqrt{2}}\bigg(a_1V_{cs}^*V_{us}
f_{K}(m_D^2-m_{K}^2)F_0^{DK}(m_{K}^2)- a_AV_{cd}^*V_{ud}
g_{1}f_Sf_D\frac{m_D^2m_{a_0}}{m_D^2-m_{a_0}^2}\bigg),
\nonumber\\
\mathcal{A}(D^+\to \pi^+\eta)&=& i\frac{G_F}{2}\bigg(\cos\phi
V_{cd}^*V_{ud} [ a_1 f_{\pi} (m_{D}^2-m_{\eta}^2)
F_0^{{D}\eta_q}(m_{\pi}^2)- 2a_A g_{1} f_S f_{D}
\frac{m_{D}^2m_{a_0}}{m_{D}^2-m_{a_0}^2}]
\nonumber\\
&&+ \sqrt{2}a_2 [V_{cd}^*V_{ud}f_{\eta}^d
+V_{cs}^*V_{us}f_{\eta}^s](m_{D}^2-m_{\pi}^2)
F_0^{{D}\pi}(m_{\eta}^2)\bigg),
\nonumber\\
\mathcal{A}(D^+\to \pi^+\eta')&=& i\frac{G_F}{2}\bigg(\sin\phi
V_{cd}^*V_{ud} [ a_1 f_{\pi} (m_{D}^2-m_{\eta'}^2)
F_0^{{D}\eta_q}(m_{\pi}^2)- 2a_A g_{1} f_S f_{D}
\frac{m_{D}^2m_{a_0}}{m_{D}^2-m_{a_0}^2}]
\nonumber\\
&&+ \sqrt{2}a_2 [V_{cd}^*V_{ud}f_{\eta'}^d
+V_{cs}^*V_{us}f_{\eta'}^s](m_{D}^2-m_{\pi}^2)
F_0^{{D}\pi}(m_{\eta'}^2)\bigg), \nonumber
\\
 \mathcal{A}(D^0\to \pi^+\pi^-)&=&
i\frac{G_F}{\sqrt{2}}V_{cd}^*V_{ud}\bigg( a_1
f_{\pi}(m_D^2-m_{\pi}^2)F_0^{D\pi}(m_{\pi}^2)- a_E
g_{1}f_Sf_D\frac{m_D^2m_{f_0(1370)}}{m_D^2-m_{f_0(1370)}^2}\bigg),
\nonumber\\
\mathcal{A}(D^0\to \pi^0\pi^0)&=&-i
\frac{G_F}{2}V_{cd}^*V_{ud}\bigg(
a_2f_{\pi}(m_D^2-m_{\pi}^2)F_0^{D\pi}(m_{\pi}^2)+a_E
g_{1}f_Sf_D\frac{m_D^2m_{f_0(1370)}}{m_D^2-m_{f_0(1370)}^2}\bigg),
\nonumber\\
\mathcal{A}(D^0\to K^+K^-)&=&i
\frac{G_F}{\sqrt{2}}V_{cs}^*V_{us}\bigg( a_1
f_{K}(m_D^2-m_{K}^2)F_0^{DK}(m_{K}^2)-a_E
g_{1}f_Sf_D\frac{m_D^2m_{f_0(1500)}}{m_D^2-m_{f_0(1500)}^2}\bigg),
\nonumber\\
\mathcal{A}(D^0\to K^0\bar K^0)&=&-i\frac{G_F}{\sqrt{2}}a_E
g_{1}f_Sf_D\bigg(V_{cs}^*V_{us}\frac{m_D^2m_{f_0(1500)}}{m_D^2-m_{f_0(1500)}^2}+V_{cd}^*V_{ud}\frac{m_D^2m_{f_0(1370)}}{m_D^2-m_{f_0(1370)}^2}\bigg)
\nonumber\\
%$D^0\to K^0\bar{K}^0$&
%$\frac{G_F}{\sqrt{2}}\bigg(V_{cs}^*V_{us}+V_{cd}^*V_{ud}\bigg)\bigg(-i
%a_A g_{KKa_0}\frac{m_D^2m_{a_0}}{m_D^2-m_{a_0}^2}f_Sf_D\bigg)$
%\\
\mathcal{A}(D^0\to \pi^0\eta)&=&i \frac{G_F}{2} \bigg(
a_2(m_{D}^2-m_{\pi}^2)
F_0^{{D}\pi}(m_{\eta}^2)[V_{cd}^*V_{ud}f_{\eta}^d
+V_{cs}^*V_{us}f_{\eta}^s]-\frac{1}{\sqrt{2}}V_{cd}^*V_{ud} a_2
f_{\pi} (m_{D}^2-m_{\eta}^2) F_0^{{D}\eta_q}(m_{\pi}^2)\cos\phi
\nonumber\\
&&+\sqrt{2}a_E
g_{1}f_Sf_D\frac{m_D^2m_{a_0}}{m_D^2-m_{a_0}^2}\cos\phi\bigg),
\nonumber\\
\mathcal{A}(D^0\to \pi^0\eta')&=& i \frac{G_F}{2} \bigg(
a_2(m_{D}^2-m_{\pi}^2)
F_0^{{D}\pi}(m_{\eta}^2)[V_{cd}^*V_{ud}f_{\eta'}^d
+V_{cs}^*V_{us}f_{\eta'}^s]-\frac{1}{\sqrt{2}}V_{cd}^*V_{ud} a_2
f_{\pi} (m_{D}^2-m_{\eta}^2) F_0^{{D}\eta_q}(m_{\pi}^2)\sin\phi
\nonumber\\
&&+\sqrt{2}a_E
g_{1}f_Sf_D\frac{m_D^2m_{a_0}}{m_D^2-m_{a_0}^2}\sin\phi\bigg),
\nonumber\\
 \mathcal{A}(D^0\to \eta\eta)&=& i
\frac{G_F}{\sqrt{2}} \bigg( a_2 (m_{D}^2-m_{\eta}^2)
F_0^{{D}\eta_q}(m_{\eta}^2)[V_{cd}^*V_{ud}f_{\eta}^d+V_{cs}^*V_{us}f_{\eta}^s]\cos\phi
\nonumber\\ &&-a_E
g_{1}f_Sf_D(\sqrt{2}V_{cs}^*V_{us}\frac{m_D^2m_{f_0(1500)}}{m_D^2-m_{f_0(1500)}^2}\sin^2\phi+
{1\over\sqrt{2}}V_{cd}^*V_{ud}\frac{m_D^2m_{f_0(1370)}}{m_D^2-m_{f_0(1370)}^2}\cos^2\phi)\bigg)
,\nonumber\\
 \mathcal{A}(D^0\to \eta\eta')&=&i
\frac{G_F}{2}\bigg( a_2(m_{D}^2-m_{\eta}^2)
F_0^{{D}\eta_q}(m_{\eta'}^2)[V_{cd}^*V_{ud}f_{\eta'}^d+V_{cs}^*V_{us}f_{\eta'}^s]\cos\phi
\nonumber\\&&+ a_2(m_{D}^2-m_{\eta'}^2)
F_0^{{D}\eta_q}(m_{\eta}^2)[V_{cd}^*V_{ud}f_{\eta}^d+V_{cs}^*V_{us}f_{\eta}^s]\sin\phi
\nonumber\\
&&+a_E
g_{1}f_Sf_D(\sqrt{2}V_{cs}^*V_{us}\frac{m_D^2m_{f_0(1500)}}{m_D^2-m_{f_0(1500)}^2}\sin2\phi-
{1\over\sqrt{2}}V_{cd}^*V_{ud}\frac{m_D^2m_{f_0(1370)}}{m_D^2-m_{f_0(1370)}^2}\sin2\phi)\bigg)
,\nonumber
\\
\mathcal{A}(D_S^+\to K^+\pi^0)&=&
-i\frac{G_F}{2}\bigg(V_{cs}^*V_{us}  a_A g_{1}f_S f_{D_S}
\frac{m_{D_S}^2m_{K^*_0}}{m_{D_S}^2-m_{K^*_0}^2} +V_{cd}^*V_{ud} a_2
f_{\pi} (m_{D_S}^2-m_K^2) F_0^{{D_S}K}(m_{\pi}^2)\bigg),
\nonumber\\
\mathcal{A}(D_S^+\to \pi^+K^0)&=&
-i\frac{G_F}{\sqrt{2}}\bigg(V_{cs}^*V_{us} a_A g_{1} f_S f_{D_S}
\frac{m_{D_S}^2m_{K^*_0}}{m_{D_S}^2-m_{K^*_0}^2}-V_{cd}^*V_{ud}a_1
f_{\pi} (m_{D_S}^2-m_K^2) F_0^{{D_S}K}(m_{\pi}^2)\bigg),
\nonumber\\
\mathcal{A}(D_S^+\to K^+\eta)&=&
-i\frac{G_F}{\sqrt{2}}\bigg(V_{cs}^*V_{us} a_A g_{1}f_S f_{D_S}
\frac{m_{D_S}^2m_{K^*_0}}{m_{D_S}^2-m_{K^*_0}^2}
(\frac{1}{\sqrt{2}}\cos\phi-\sin\phi)+ V_{cs}^*V_{us}a_1 f_{K}
(m_{D}^2-m_{\eta}^2) F_0^{{D_S}\eta_s}(m_{K}^2)\sin\phi
\nonumber\\
&&- a_2 (m_{D}^2-m_{K}^2)
F_0^{{D}K}(m_{\eta}^2)[V_{cd}^*V_{ud}f_{\eta}^d
+V_{cs}^*V_{us}f_{\eta}^s]\bigg),
\nonumber\\
\mathcal{A}(D_S^+\to
K^+\eta')&=&-i\frac{G_F}{\sqrt{2}}\bigg(V_{cs}^*V_{us} a_A g_{1}f_S
f_{D_S} \frac{m_{D_S}^2m_{K^*_0}}{m_{D_S}^2-m_{K^*_0}^2}
(\frac{1}{\sqrt{2}}\sin\phi+\cos\phi)-V_{cs}^*V_{us}a_1 f_{K}
(m_{D}^2-m_{\eta'}^2) F_0^{{D_S}\eta_s}(m_{K}^2)\cos\phi
\nonumber\\
&&- a_2 (m_{D}^2-m_{K}^2)
F_0^{{D}K}(m_{\eta'}^2)[V_{cd}^*V_{ud}f_{\eta'}^d
+V_{cs}^*V_{us}f_{\eta'}^s]\bigg),
\end{eqnarray}}
 The formulas for doubly Cabibbo-suppressed decays of $D\to PP$
are listed as {
\begin{eqnarray}
\mathcal{A}(D^+\to K^+\pi^0)&=& -i\frac{G_F}{2}V_{cd}^*V_{us} \bigg(
a_A g_{1} f_S f_D\frac{m_D^2m_{K^*_0}}{m_D^2-m_{K^*_0}^2} + a_1
f_{K} (m_D^2-m_{\pi}^2) F_0^{D\pi}(m_{K}^2)\bigg),
\nonumber\\
\mathcal{A}(D^+\to \pi^+K^0)&=& -i\frac{G_F}{\sqrt{2}}V_{cd}^*V_{us}
\bigg( a_A g_{1}f_S f_D \frac{m_D^2m_{K^*_0}}{m_D^2-m_{K^*_0}^2} -
a_2 f_{K} (m_D^2-m_{\pi}^2) F_0^{D\pi}(m_{K}^2)\bigg),
\nonumber\\
\mathcal{A}(D^+\to K^+\eta)&=& i\frac{G_F}{2}V_{cd}^*V_{us} \bigg(
a_A g_{1} f_S f_{D}\frac{m_{D}^2m_{K^*_0}}{m_{D}^2-m_{K^*_0}^2}
(\sqrt{2}\sin\phi-\cos\phi) + a_1 f_{K} (m_{D}^2-m_{\eta}^2)
F_0^{{D}\eta_q}(m_{K}^2)\cos\phi\bigg),
\nonumber\\
\mathcal{A}(D^+\to K^+\eta')&=& i\frac{G_F}{2}V_{cd}^*V_{us} \bigg(-
a_A g_{1}f_S f_{D} \frac{m_{D}^2m_{K^*_0}}{m_{D}^2-m_{K^*_0}^2}
(\sin\phi+\sqrt{2}\cos\phi) + a_1 f_{K} (m_{D}^2-m_{\eta'}^2)
F_0^{{D}\eta_q}(m_{K}^2)\sin\phi\bigg), \nonumber\\
\mathcal{A}(D^0\to K^+\pi^-)&=& -i\frac{G_F}{\sqrt{2}}V_{cd}^*V_{us}
\bigg( a_E g_{1}f_S f_D \frac{m_D^2m_{K^*_0}}{m_D^2-m_{K^*_0}^2} -
a_1 f_{K} (m_D^2-m_{\pi}^2) F_0^{D\pi}(m_{K}^2)\bigg),
\nonumber\\
\mathcal{A}(D^0\to K^0\pi^0)&=& i\frac{G_F}{2}V_{cd}^*V_{us} \bigg(
a_E g_{1} f_S f_D \frac{m_D^2m_{K^*_0}}{m_D^2-m_{K^*_0}^2}+ a_2
f_{K} (m_D^2-m_{\pi}^2) F_0^{D\pi}(m_{K}^2)\bigg),
\nonumber\\
\mathcal{A}(D^0\to K^0\eta)&=&i \frac{G_F}{2}V_{cd}^*V_{us}\bigg(a_E
g_{1}f_Sf_{D}\frac{m_D^2m_{K^*_0}}{m_D^2-m_{K^*_0}^2}(\sqrt{2}\sin\phi-\cos\phi)
+a_2f_K(m_D^2-m_{\eta}^2) F_0^{D\eta_q}(m_K^2)\cos{\phi}\bigg),
\nonumber\\
\mathcal{A}(D^0\to K^0\eta')&=&i
\frac{G_F}{2}V_{cd}^*V_{us}\bigg(-a_E
g_{1}f_Sf_{D}\frac{m_D^2m_{K^*_0}}{m_D^2-m_{K^*_0}^2}(\sqrt{2}\cos\phi+\sin\phi)
+a_2f_K(m_D^2-m_{\eta'}^2) F_0^{D\eta_q}(m_K^2)\sin{\phi}\bigg),
\nonumber\\
\mathcal{A}(D_s^+\to K^0K^+)&=& i\frac{G_F}{\sqrt{2}}V_{cd}^*V_{us}
 f_{K} (m_{D_s}^2-m_{K}^2) F_0^{D_s K}(m_{K}^2)(a_1+a_2).
\end{eqnarray}
}

{The formulas for Cabibbo-favored decays of $D\to PV$ are shown as }
\begin{eqnarray}
\mathcal{A}(D^0\to \pi^+K^{*-})&=&
\sqrt{2}G_FV_{cs}^*V_{ud}\bigg(a_E^{PV}g_{s}f_{K}f_{D}\frac{m_{D}^2}{m_{D}^2-m_{K}^2}+a_1^{PV}m_{K^*}f_{\pi}A_0^{DK^*}(m_{\pi}^2)
\bigg)(\varepsilon^*\cdot p_{D}),\nonumber\\
\mathcal{A}(D^0\to K^-\rho^+)&=&
\sqrt{2}G_FV_{cs}^*V_{ud}\bigg(a_E^{PV}g_{s}f_{K}f_{D}\frac{m_{D}^2}{m_{D}^2-m_{K}^2}+a_1^{PV}m_{\rho}f_{\rho}F_1^{D
K}(m_{\rho}^2) \bigg)(\varepsilon^*\cdot p_{D}),\nonumber\\
\mathcal{A}(D^0\to \pi^0\bar{K}^{*0})&=&
G_FV_{cs}^*V_{ud}\bigg(-a_E^{PV}g_{s}f_{K}f_{D}\frac{m_{D}^2}{m_{D}^2-m_{K}^2}+a_2^{PV}m_{K^*}f_{K^*}
F_1^{D\pi}(m_{K^*}^2) \bigg)(\varepsilon^*\cdot
p_{D}),\nonumber\\
 \mathcal{A}(D^0\to \bar{K}^0\rho^0)&=&
G_FV_{cs}^*V_{ud}\bigg(-a_E^{PV}g_{s}f_{K}f_{D}\frac{m_{D}^2}{m_{D}^2-m_{K}^2}+a_2^{PV}m_{\rho}f_{K}A_0^{D\rho}(m_{K}^2)
\bigg)(\varepsilon^*\cdot p_{D}),\nonumber\\
 \mathcal{A}(D^0\to\eta\bar{K}^{*0})&=&
G_FV_{cs}^*V_{ud}\bigg(a_E^{PV}f_{K}f_{D}\frac{m_{D}^2}{m_{D}^2-m_{K}^2}(g_{s}\cos\phi-\sqrt{2}g_{ss}\sin\phi)+a_2^{PV}m_{K^*}f_{K^*}
F_1^{D\eta_q}(m_{K^*}^2)\cos\phi
\bigg)(\varepsilon^*\cdot p_{D}),\nonumber\\
 \mathcal{A}(D^0\to
\eta'\bar{K}^{*0})&=&
G_FV_{cs}^*V_{ud}\bigg(a_E^{PV}f_{K}f_{D}\frac{m_{D}^2}{m_{D}^2-m_{K}^2}(g_{s}\sin\phi+\sqrt{2}g_{ss}\cos\phi)+a_2^{PV}m_{K^*}f_{K^*}
F_1^{D\eta_q}(m_{K^*}^2)\sin\phi
\bigg)(\varepsilon^*\cdot p_{D}),\nonumber\\
 \mathcal{A}(D^0\to
\bar{K}^0\omega)&=&
G_FV_{cs}^*V_{ud}\bigg(a_E^{PV}g_{s}f_{K}f_{D}\frac{m_{D}^2}{m_{D}^2-m_{K}^2}+a_2^{PV}m_{\omega}f_{K}A_0^{D\omega}(m_{K}^2)
\bigg)(\varepsilon^*\cdot p_{D}),\nonumber\\
 \mathcal{A}(D^0\to
\bar{K}^0\phi)&=&
\sqrt{2}G_FV_{cs}^*V_{ud}a_E^{PV}g_{ss}f_{K}f_{D}\frac{m_{D}^2}{m_{D}^2-m_{K}^2}
(\varepsilon^*\cdot p_{D}),\nonumber
\\
 \mathcal{A}(D^+\to
\pi^+\bar{K}^{*0})&=&
\sqrt{2}G_FV_{cs}^*V_{ud}m_{K^*}\bigg(a_1^{PV}f_{\pi}A_0^{DK^*}(m_{\pi}^2)+a_2^{PV}f_{K^*}F_1^{D\pi}(m_{K^*}^2)\bigg)(\varepsilon^*\cdot
p_{D}),\nonumber\\
 \mathcal{A}(D^+\to \bar{K}^0\rho^+)&=&
\sqrt{2}G_FV_{cs}^*V_{ud}m_{\rho}\bigg(a_2^{PV}f_{K}A_0^{D{\rho}}(m_{K}^2)+a_1^{PV}f_{{\rho}}F_1^{DK}(m_{\rho}^2)\bigg)(\varepsilon^*\cdot
p_{D}),\nonumber
\\
 \mathcal{A}(D_S^+\to K^+\bar{K}^{*0})&=&
\sqrt{2}G_FV_{cs}^*V_{ud}\bigg(a_A^{PV}g_{s}f_{\pi}f_{D_S}\frac{m_{D_S}^2}{m_{D_S}^2-m_{\pi}^2}+a_2^{PV}m_{K^*}f_{K^*}F_1^{{D_S}K}(m_{K^*}^2)
\bigg)(\varepsilon^*\cdot p_{D_S}),\nonumber\\
 \mathcal{A}(D_S^+\to
\bar{K}^0K^{*+})&=&
\sqrt{2}G_FV_{cs}^*V_{ud}\bigg(a_A^{PV}g_{s}f_{\pi}f_{D_S}\frac{m_{D_S}^2}{m_{D_S}^2-m_{\pi}^2}+a_2^{PV}m_{K^*}f_{K}A_0^{{D_S}K^*}(m_{K}^2)
\bigg)(\varepsilon^*\cdot p_{D_S}),\nonumber
\end{eqnarray}
\begin{eqnarray}
 \mathcal{A}(D_S^+\to
\eta\rho^+)&=&-
\sqrt{2}G_FV_{cs}^*V_{ud}a_1^{PV}m_{\rho}f_{\rho}F_1^{{D_S}\eta_s}(m_{\rho}^2)\sin\phi
(\varepsilon^*\cdot p_{D_S}),\nonumber\\
 \mathcal{A}(D_S^+\to
\eta'\rho^+)&=&
\sqrt{2}G_FV_{cs}^*V_{ud}a_1^{PV}m_{\rho}f_{\rho}F_1^{{D_S}\eta_s}(m_{\rho}^2)\cos\phi
(\varepsilon^*\cdot p_{D_S}),\nonumber
\\
 \mathcal{A}(D_S^+\to
\pi^+\phi)&=&
\sqrt{2}G_FV_{cs}^*V_{ud}a_1^{PV}m_{\phi}f_{\pi}A_0^{{D_S}\phi}(m_{\pi}^2)
(\varepsilon^*\cdot p_{D_S}),\nonumber
\\
\mathcal{A}(D_S^+\to \pi^+\rho^0) & =& 2G_FV_{cs}^*V_{ud}a_A^{PV}
g_q f_{\pi}f_{D_S}
\frac{m_{D_S}^2}{m_{D_S}^2-m_{\pi}^2}(\varepsilon^*\cdot p_{D_S}),
\nonumber
\\
 \mathcal{A}( D_S^+\to
\pi^0\rho^+)&=&2G_FV_{cs}^*V_{ud}a_A^{PV} g_q f_{\pi}f_{D_S}
\frac{m_{D_S}^2}{m_{D_S}^2-m_{\pi}^2}(\varepsilon^*\cdot p_{D_S}),
\nonumber
\\
\mathcal{A}( D_S^+\to \pi^+\omega)&=&0,
\end{eqnarray}
 where the
effective strong coupling constants between  the intermediate state
and the  final states for $D\to PV$ are $g_q=4.2$, $g_s=4.6$, and
$g_{ss}=4.5$.
 The formulas
for singly Cabibbo-suppressed decays of $D\to PV$ are shown as

{
\begin{eqnarray}
\mathcal{A}(D^0\to \pi^+\rho^-)&=&-
G_FV_{cd}^*V_{ud}f_{\pi}\bigg(a_E^{PV}g_{q}f_{D}
\frac{m_{D}^2}{m_{D}^2-m_{\pi}^2}-\sqrt{2}a_1^{PV}m_{\rho}A_0^{D\rho}(m_{\pi}^2)
\bigg)(\varepsilon^*\cdot p_{D}),
\nonumber\\
\mathcal{A}(D^0\to \pi^-\rho^+)&=&-
G_FV_{cd}^*V_{ud}\bigg(a_E^{PV}g_{q}f_{\pi}f_{D}\frac{m_{D}^2}{m_{D}^2-m_{\pi}^2}-\sqrt{2}a_1^{PV}m_{\rho}
f_{\rho}F_1^{D \pi}(m_{\rho}^2) \bigg)(\varepsilon^*\cdot p_{D}),
\nonumber\\
\mathcal{A}(D^0\to \pi^0\rho^0)&=&
-\frac{G_F}{\sqrt{2}}V_{cd}^*V_{ud}a_2^{PV}m_{\rho}\bigg(
f_{\rho}F_1^{D\pi}(m_{\rho}^2) +f_{\pi}A_0^{D \rho}(m_{\pi}^2)
\bigg)(\varepsilon^*\cdot p_{D}),\nonumber
\\
 \mathcal{A}(D^0\to K^-K^{*+})&=&
\sqrt{2}G_FV_{cs}^*V_{us}\bigg(a_E^{PV}g_{ss}f_{D}m_{D}^2(\frac{f_{\eta}^s}{m_{D}^2-m_{\eta}^2}+\frac{f_{\eta'}^s}{m_{D}^2-m_{\eta'}^2})+a_1^{PV}m_{K^*}
f_{K^*}F_1^{D {K}}(m_{K^*}^2) \bigg)(\varepsilon^*\cdot p_{D}),
\nonumber\\
\mathcal{A}(D^0\to K^+K^{*-})&=&
\sqrt{2}G_FV_{cs}^*V_{us}\bigg(a_E^{PV}g_{ss}f_{D}m_{D}^2(\frac{f_{\eta}^s}{m_{D}^2-m_{\eta}^2}+\frac{f_{\eta'}^s}{m_{D}^2-m_{\eta'}^2})+a_1^{PV}m_{K^*}
f_{K}A_0^{D {K^*}}(m_{K}^2) \bigg)(\varepsilon^*\cdot p_{D}),
\nonumber\\
\mathcal{A}(D^0\to \bar{K}^0K^{*0})&=&
\sqrt{2}G_Fa_E^{PV}f_{D}m_{D}^2\bigg(V_{cd}^*V_{ud}g_{s}\frac{f_{\pi}}{m_{D}^2-m_{\pi}^2}+V_{cs}^*V_{us}g_{ss}
[\frac{f_{\eta}^s}{m_{D}^2-m_{\eta}^2}+\frac{f_{\eta'}^s}{m_{D}^2-m_{\eta'}^2}]\bigg)(\varepsilon^*\cdot
p_{D}),
\nonumber\\
\mathcal{A}(D^0\to K^0\bar{K}^{*0})&=&
\sqrt{2}G_Fa_E^{PV}f_{D}m_{D}^2\bigg(V_{cd}^*V_{ud}g_{s}\frac{f_{\pi}}{m_{D}^2-m_{\pi}^2}+V_{cs}^*V_{us}g_{ss}
[\frac{f_{\eta}^s}{m_{D}^2-m_{\eta}^2}+\frac{f_{\eta'}^s}{m_{D}^2-m_{\eta'}^2}]\bigg)(\varepsilon^*\cdot
p_{D}),
\nonumber\\
\mathcal{A}(D^0\to \pi^0\omega)&=&
\frac{G_F}{\sqrt{2}}V_{cd}^*V_{ud}a_2^{PV}m_{\omega}\bigg(f_{\omega}F_1^{D
\pi}(m_{\omega}^2)-f_{\pi} A_0^{D {\omega}}(m_{\pi}^2)
\bigg)(\varepsilon^*\cdot p_{D}),
\nonumber\\
\mathcal{A}(D^0\to \pi^0\phi)&=&
G_FV_{cs}^*V_{us}a_2^{PV}m_{\phi}f_{\phi}F_1^{D
\pi}(m_{\phi}^2)(\varepsilon^*\cdot p_{D})
\nonumber\\
\mathcal{A}(D^0\to
\eta\phi)&=&G_FV_{cs}^*V_{us}a_2^{PV}m_{\phi}f_{\phi}F_1^{D
\eta_q}(m_{\phi}^2)\cos\phi(\varepsilon^*\cdot p_{D}),
\nonumber\\
\mathcal{A}(D^0\to \eta\rho^0)&=&
G_Fa_2^{PV}m_{\rho}\bigg([V_{cd}^*V_{ud}f_{\eta}^{d}+V_{cs}^*V_{us}f_{\eta}^{s}]A_0^{D
\rho}(m_{\eta}^2) -\frac{1}{\sqrt{2}}V_{cd}^*V_{ud}f_{\rho}F_1^{D
{\eta_q}}(m_{\rho}^2)\cos\phi \bigg)(\varepsilon^*\cdot p_{D}),
\nonumber\\
\mathcal{A}(D^0\to \eta'\rho^0)&=& G_Fa_2^{PV}
m_{\rho}\bigg([V_{cd}^*V_{ud}f_{\eta'}^{d}+V_{cs}^*V_{us}f_{\eta'}^{s}]A_0^{D
\rho}(m_{\eta'}^2) -\frac{1}{\sqrt{2}}V_{cd}^*V_{ud}f_{\rho}F_1^{D
{\eta_q}}(m_{\rho}^2)\sin\phi \bigg)(\varepsilon^*\cdot p_{D})
\nonumber\\
\mathcal{A}(D^0\to \eta\omega)&=&{G_F}
a_2^{PV}m_{\omega}\bigg([V_{cd}^*V_{ud}f_{\eta}^{d}+V_{cs}^*V_{us}f_{\eta}^{s}]A_0^{D
\omega}(m_{\eta'}^2) +\frac{1}{\sqrt{2}}\cos\phi
V_{cd}^*V_{ud}f_{\omega}F_1^{D {\eta_q}}(m_{\omega}^2) \bigg)
(\varepsilon^*\cdot p_{D}) ,
\nonumber\\
\mathcal{A}(D^0\to \eta'\omega)&=&  {G_F}
a_2^{PV}m_{\omega}\bigg([V_{cd}^*V_{ud}f_{\eta'}^{d}+V_{cs}^*V_{us}f_{\eta'}^{s}]A_0^{D
\omega}(m_{\eta'}^2) +\frac{1}{\sqrt{2}}\sin\phi
V_{cd}^*V_{ud}f_{\omega}F_1^{D {\eta_q}}(m_{\omega}^2) \bigg)
(\varepsilon^*\cdot p_{D}) ,\nonumber
\\
 \mathcal{A}(D^+\to
\pi^+\rho^0)&=& {G_F}V_{cd}^*V_{ud}m_{\rho}\bigg(2a_A^{PV} g_q
f_{\pi}f_{D}
\frac{m_{D}^2}{m_{D}^2-m_{\pi}^2}-a_2^{PV}f_{\rho}F_1^{D
\pi}(m_{\rho}^2)-a_1^{PV}f_{\pi}A_0^{D \rho} (m_{\pi}^2)
\bigg)(\varepsilon^*\cdot p_{D}),
\nonumber\\
\mathcal{A}(D^+\to \pi^0\rho^+)&=&
{G_F}V_{cd}^*V_{ud}m_{\rho}\bigg(2a_A^{PV} g_q f_{\pi}f_{D}
\frac{m_{D}^2}{m_{D}^2-m_{\pi}^2}-a_1^{PV}f_{\rho}F_1^{D
\pi}(m_{\rho}^2)-a_2^{PV}f_{\pi} A_0^{D \rho}(m_{\pi}^2)
\bigg)(\varepsilon^*\cdot p_{D}),
\nonumber\\
\mathcal{A}(D^+\to K^+\bar{K}^{*0})&=&
\sqrt{2}{G_F}\bigg(V_{cd}^*V_{ud}a_A^{PV}g_{s}f_{\pi}f_{D}\frac{m_{D}^2}{m_{D}^2-m_{\pi}^2}+V_{cs}^*V_{us}a_1^{PV}m_{K^*}
f_{K}A_0^{D {K^*}}(m_{K}^2) \bigg)(\varepsilon^*\cdot p_{D}),
\nonumber\\
\mathcal{A}(D^+\to \bar{K}^0K^{*+})&=&
\sqrt{2}{G_F}\bigg(V_{cd}^*V_{ud}a_A^{PV}g_{s}f_{\pi}f_{D}\frac{m_{D}^2}{m_{D}^2-m_{\pi}^2}+V_{cs}^*V_{us}a_1^{PV}m_{K^*}
f_{K^*}F_1^{D {K}}(m_{K^*}^2) \bigg)(\varepsilon^*\cdot p_{D})
\nonumber
\end{eqnarray}
\begin{eqnarray}
 \mathcal{A}(D^+\to \eta\rho^+)&=&
{G_F}m_{\rho}\bigg(\sqrt{2}a_2^{PV}A_0^{D
\rho}(m_{\eta}^2)[V_{cd}^*V_{ud}f_{\eta}^{d}+V_{cs}^*V_{us}f_{\eta}^{s}]
+V_{cd}^*V_{ud}a_1^{PV}f_{\rho} F_1^{D {\eta_q}}(m_{\rho}^2)\cos\phi
\bigg)(\varepsilon^*\cdot p_{D}),
\nonumber\\
\mathcal{A}(D^+\to \eta'\rho^+)&=&
{G_F}m_{\rho}\bigg(\sqrt{2}a_2^{PV}A_0^{D
\rho}(m_{\eta'}^2)[V_{cd}^*V_{ud}f_{\eta'}^{d}+V_{cs}^*V_{us}f_{\eta'}^{s}]
+V_{cd}^*V_{ud}a_1^{PV}f_{\rho} F_1^{D {\eta_q}}(m_{\rho}^2)\sin\phi
\bigg)(\varepsilon^*\cdot p_{D}),
\nonumber\\
\mathcal{A}(D^+\to \pi^+\phi)&=&
\sqrt{2}{G_F}{}V_{cs}^*V_{us}a_2^{PV}m_{\phi}f_{\phi}F_1^{D
\pi}(m_{\phi}^2) (\varepsilon^*\cdot p_{D}),
\nonumber\\
\mathcal{A}(D^+\to \pi^+\omega)&=&
{G_F}{}V_{cd}^*V_{ud}m_{\omega}\bigg(a_2^{PV}f_{\omega}F_1^{D
\pi}(m_{\omega}^2)+a_1^{PV}f_{\pi} A_0^{D {\omega}}(m_{\pi}^2)
\bigg)(\varepsilon^*\cdot p_{D}) ,\nonumber
\\
\mathcal{A}(D_S^+\to\pi^+K^{*0})&=&
\sqrt{2}{G_F}{}\bigg(V_{cs}^*V_{us}a_A^{PV}g_{s}f_{K}f_{D_S}\frac{m_{D_S}^2}{m_{D_S}^2-m_{K}^2}+V_{cd}^*V_{ud}a_1^{PV}m_{K^*}
f_{\pi}A_0^{D_S {K^*}}(m_{\pi}^2) \bigg)(\varepsilon^*\cdot
p_{D_S}),
\nonumber\\
\mathcal{A}(D_S^+\to \pi^0K^{*+})&=&
{G_F}\bigg(V_{cs}^*V_{us}a_A^{PV}g_{s}f_{K}f_{D_S}\frac{m_{D_S}^2}{m_{D_S}^2-m_{K}^2}
-V_{cd}^*V_{ud}a_2^{PV}m_{K^*}f_{\pi} A_0^{D_S {K^*}}(m_{\pi}^2)
\bigg)(\varepsilon^*\cdot p_{D_S}),
\nonumber\\
\mathcal{A}(D_S^+\to K^+\rho^0)&=&
{G_F}\bigg(V_{cs}^*V_{us}a_A^{PV}g_{s}f_{K}f_{D_S}\frac{m_{D_S}^2}{m_{D_S}^2-m_{K}^2}-V_{cd}^*V_{ud}a_2^{PV}m_{\rho}f_{\rho}
F_1^{D_S {K}}(m_{\rho}^2) \bigg)(\varepsilon^*\cdot p_{D_S}),
\nonumber\\
\mathcal{A}(D_S^+\to K^0\rho^+)&=&
\sqrt{2}{G_F}{}\bigg(V_{cs}^*V_{us}a_A^{PV}g_{s}f_{K}f_{D_S}\frac{m_{D_S}^2}{m_{D_S}^2-m_{K}^2}+V_{cd}^*V_{ud}a_1^{PV}m_{\rho}
f_{\rho}F_1^{D_S {K}}(m_{\rho}^2) \bigg)(\varepsilon^*\cdot
p_{D_S}), \nonumber
\\
 \mathcal{A}(D_S^+\to\eta K^{*+})&=&
{G_F}{}\Bigg(V_{cs}^*V_{us}[a_A^{PV}f_{K}f_{D_S}\frac{m_{D_S}^2}{m_{D_S}^2-m_{K}^2}
(g_{s}\cos\phi-\sqrt{2}g_{ss}\sin\phi)
-\sqrt{2}a_1^{PV}m_{K^*}f_{K^*}F_1^{D_S \eta_s}\sin\phi]
\nonumber\\
&&+\sqrt{2}a_2^{PV}m_{K^*}A_0^{D_S
{K^*}}(m_{\eta}^2)[V_{cd}^*V_{ud}f_{\eta}^d+V_{cs}^*V_{us}f_{\eta}^s]\Bigg)(\varepsilon^*\cdot
p_{D_S}),
\nonumber\\
\mathcal{A}(D_S^+\to\eta' K^{*+})&=&
{G_F}{}\Bigg(V_{cs}^*V_{us}[a_A^{PV}f_{K}f_{D_S}\frac{m_{D_S}^2}{m_{D_S}^2-m_{K}^2}
(g_{s}\sin\phi+\sqrt{2}g_{ss}\cos\phi)
+\sqrt{2}a_1^{PV}m_{K^*}f_{K^*}F_1^{D_S \eta_s}\cos\phi]
\nonumber\\
&&+\sqrt{2}a_2^{PV}m_{K^*}A_0^{D_S
{K^*}}(m_{\eta'}^2)[V_{cd}^*V_{ud}f_{\eta'}^d+V_{cs}^*V_{us}f_{\eta'}^s]\Bigg)(\varepsilon^*\cdot
p_{D_S}),
\nonumber\\
\mathcal{A}(D_S^+\to K^+\phi)&=&
\sqrt{2}{G_F}{}V_{cs}^*V_{us}\bigg(a_A^{PV}g_{ss}f_{K}f_{D_S}\frac{m_{D_S}^2}{m_{D_S}^2-m_{K}^2}+a_2^{PV}m_{\phi}f_{\phi}
F_1^{D_S {K}}(m_{\phi}^2)+a_1^{PV}m_{\phi}f_{K}A_0^{D_S \phi}(m_K^2)
\bigg)(\varepsilon^*\cdot p_{D_S}) , \nonumber\\
\mathcal{A}(D_S^+\to K^+\omega)&=&
{G_F}{}\bigg(V_{cs}^*V_{us}a_A^{PV}g_{s}f_{K}f_{D_S}\frac{m_{D_S}^2}{m_{D_S}^2-m_{K}^2}+V_{cd}^*V_{ud}
a_2^{PV}m_{\omega}f_{\omega}F_1^{D_S {K}}(m_{\omega}^2)
\bigg)(\varepsilon^*\cdot p_{D_S}).
\end{eqnarray}
}

{\normalsize The formulas for doubly Cabibbo-suppressed decays of
$D\to PV$ are shown as}
\begin{eqnarray}
\mathcal{A}(D^0\to \pi^-K^{*+})&=&
\sqrt{2}{G_F}{}V_{cd}^*V_{us}\bigg(a_E^{PV}g_{s}f_{K}f_{D}\frac{m_{D}^2}{m_{D}^2-m_{K}^2}+a_1^{PV}m_{K^*}f_{K^*}F_1^{D
\pi}(m_{K^*}^2) \bigg)(\varepsilon^*\cdot p_{D}),
\nonumber\\
\mathcal{A}(D^0\to \pi^0K^{*0})&=&
{G_F}{}V_{cd}^*V_{us}\bigg(a_2^{PV}m_{K^*} f_{K^*}F_1^{D
\pi}(m_{K^*}^2)-a_E^{PV}g_{s}f_{K}f_{D}\frac{m_{D}^2}{m_{D}^2-m_{K}^2}
\bigg)(\varepsilon^*\cdot p_{D}),
\nonumber\\
\mathcal{A}(D^0\to K^+\rho^-)&=&
\sqrt{2}{G_F}{}V_{cd}^*V_{us}f_{K}\bigg(a_1^{PV}m_{\rho}A_0^{D
\rho}(m_{K}^2)-a_E^{PV}g_{s}f_{D}\frac{m_{D}^2}{m_{D}^2-m_{K}^2}
\bigg)(\varepsilon^*\cdot p_{D}),
\nonumber\\
\mathcal{A}(D^0\to K^0\rho^0)&=&
{G_F}{}V_{cd}^*V_{us}f_{K}\bigg(a_2^{PV}m_{\rho}A_0^{D
\rho}(m_{K}^2)-a_E^{PV}g_{s}f_{D}\frac{m_{D}^2}{m_{D}^2-m_{K}^2}
\bigg)(\varepsilon^*\cdot p_{D}),
\nonumber\\
\mathcal{A}(D^0\to K^0\omega)&=&
{G_F}V_{cd}^*V_{us}f_{K}\bigg(a_2^{PV}m_{\omega}A_0^{D
\omega}(m_{K}^2)-a_E^{PV}g_{s}f_{D}\frac{m_{D}^2}{m_{D}^2-m_{K}^2}\bigg)(\varepsilon^*\cdot
p_{D}),
\nonumber\\
\mathcal{A}(D^0\to K^0\phi)&=&
\sqrt{2}{G_F}{}V_{cd}^*V_{us}a_E^{PV}g_{s}f_{K}f_{D}\frac{m_{D}^2}{m_{D}^2-m_{K}^2}(\varepsilon^*\cdot
p_{D}),
\nonumber\\
\mathcal{A}(D^0\to\eta K^{*0})&=&
{G_F}V_{cd}^*V_{us}\bigg(a_E^{PV}f_{K}f_{D}\frac{m_{D}^2}{m_{D}^2-m_{K}^2}(g_{s}\cos\phi-\sqrt{2}g_{ss}\sin\phi)+a_2^{PV}m_{K^*}f_{K^*}F_1^{D
\eta_q}(m_{K^*}^2)\cos\phi \bigg)(\varepsilon^*\cdot p_{D}),
\nonumber\\
\mathcal{A}(D^0\to\eta' K^{*0})&=&
{G_F}V_{cd}^*V_{us}\bigg(a_E^{PV}f_{K}f_{D}\frac{m_{D}^2}{m_{D}^2-m_{K}^2}(g_{s}\sin\phi+\sqrt{2}g_{ss}\cos\phi)+a_2^{PV}m_{K^*}f_{K^*}F_1^{D
\eta_q}(m_{K^*}^2)\sin\phi \bigg)(\varepsilon^*\cdot p_{D})
,\nonumber
\end{eqnarray}
\begin{eqnarray}
 \mathcal{A}(D^+\to \pi^+K^{*0})&=&
\sqrt{2}{G_F}{}V_{cd}^*V_{us}\bigg(a_A^{PV}g_{s}f_{K}f_{D}\frac{m_{D}^2}{m_{D}^2-m_{K}^2}+a_2^{PV}m_{K^*}f_{K^*}F_1^{D
\pi}(m_{K^*}^2) \bigg)(\varepsilon^*\cdot p_{D}),
\nonumber\\
\mathcal{A}(D^+\to \pi^0K^{*+})&=&
{G_F}V_{cd}^*V_{us}\bigg(a_A^{PV}g_{s}f_{K}f_{D}\frac{m_{D}^2}{m_{D}^2-m_{K}^2}-a_1^{PV}m_{K^*}f_{K^*}F_1^{D
\pi}(m_{K^*}^2)\bigg)(\varepsilon^*\cdot p_{D}),
\nonumber\\
\mathcal{A}(D^+\to K^+\rho^0)&=&
{G_F}V_{cd}^*V_{us}f_{K}\bigg(a_A^{PV}g_{s}f_{D}\frac{m_{D}^2}{m_{D}^2-m_{K}^2}-a_1^{PV}m_{\rho}A_0^{D
\rho}(m_{K}^2) \bigg)(\varepsilon^*\cdot p_{D}),
\nonumber\\
\mathcal{A}(D^+\to K^0\rho^+)&=&
\sqrt{2}{G_F}{}V_{cd}^*V_{us}f_{K}\bigg(a_A^{PV}g_{s}f_{D}\frac{m_{D}^2}{m_{D}^2-m_{K}^2}
+a_2^{PV}m_{\rho}A_0^{D \rho}(m_{K}^2) \bigg)(\varepsilon^*\cdot
p_{D}),
\nonumber\\
\mathcal{A}(D^+\to K^+\omega)&=&
{G_F}V_{cd}^*V_{us}f_{K}\bigg(a_A^{PV}g_{s}f_{D}\frac{m_{D}^2}{m_{D}^2-m_{K}^2}
+a_1^{PV}m_{\omega}A_0^{D\omega}(m_{K}^2) \bigg)(\varepsilon^*\cdot
p_{D}),
\nonumber\\
\mathcal{A}(D^+\to K^+\phi)&=&
\sqrt{2}{G_F}{}V_{cd}^*V_{us}a_A^{PV}g_{ss}f_{K}f_{D}\frac{m_{D}^2}{m_{D}^2-m_{K}^2}
(\varepsilon^*\cdot p_{D}), \nonumber
\\
\mathcal{A}(D^+\to\eta K^{*+})&=&
{G_F}{}V_{cd}^*V_{us}\bigg(a_A^{PV}f_{K}f_{D}\frac{m_{D}^2}{m_{D}^2-m_{K}^2}(g_{s}\cos\phi-\sqrt{2}g_{ss}\sin\phi)+a_1^{PV}m_{K^*}f_{K^*}F_1^{D
\eta_q}(m_{K^*}^2)\cos\phi \bigg)(\varepsilon^*\cdot p_{D}),
\nonumber\\
\mathcal{A}(D^+\to\eta' K^{*+})&=&
{G_F}V_{cd}^*V_{us}\bigg(a_A^{PV}f_{K}f_{D}\frac{m_{D}^2}{m_{D}^2-m_{K}^2}(g_{s}\sin\phi+\sqrt{2}g_{ss}\cos\phi)+a_1^{PV}m_{K^*}f_{K^*}F_1^{D
\eta_q}(m_{K^*}^2)\sin\phi \bigg)(\varepsilon^*\cdot p_{D}),
\nonumber\\
 \mathcal{A}(D_S^+\to
K^+K^{*0})&=&
\sqrt{2}{G_F}V_{cd}^*V_{us}m_{K^*}\bigg(a_2
f_{K^*}F_1^{{D_S}
{K}}(m_{K^*}^2)+a_1^{PV}f_{K}A_0^{{D_S} {K^*}}
(m_{K}^2)\bigg)(\varepsilon^*\cdot p_{D}),
\nonumber\\
\mathcal{A}(D_S^+\to K^0K^{*+})&=&
\sqrt{2}{G_F}V_{cd}^*V_{us}m_{K^*}\bigg(a_1^{PV}f_{K^*}F_1^{{D_S}
{K}}(m_{K^*}^2)+a_2^{PV}f_{K}A_0^{{D_S} {K^*}}
(m_{K}^2)\bigg)(\varepsilon^*\cdot p_{D}).
\end{eqnarray}

%%%%%%%%%%%%%%%%%%%%%%%%%%%%%%%%%%%%%%%%%%%%%%%%%%%%%%%%%%%%%%%%%%%%%%%%%%%%%%%%%%%%%%%%%%%%%%%%%%%%%%%%%%%%%%%%%%%%%%%%%%%%%%%%%%%%%%%%%%%%%%%%%%%%

%\section{REFERENCES }

\end{document}